\pdfoutput=1

\documentclass[pdflatex,sn-basic]{sn-jnl}
\usepackage{natbib}
   \setcitestyle{aysep={}}

\usepackage{graphicx}%
\usepackage{multirow}%
\usepackage{amsmath,amssymb,amsfonts}%
\usepackage{amsthm}%
\usepackage{mathrsfs}%
\usepackage[title]{appendix}%
\usepackage{xcolor}%
\usepackage{textcomp}%
\usepackage{manyfoot}%
\usepackage{booktabs}%
\usepackage{algorithm}%
\usepackage{algorithmicx}%
\usepackage{algpseudocode}%
\usepackage{listings}%
\usepackage{caption}
\usepackage{subcaption}
\usepackage{graphicx,epstopdf}
\usepackage{rotating}

\begin{document}

\title[A finite mixture approach for the analysis of digital skills in Bulgaria, Finland and Italy: the role of socio-economic factors]{A finite mixture approach for the analysis of digital skills in Bulgaria, Finland and Italy: the role of socio-economic factors}


\author*[1]{\fnm{Dalila} \sur{Failli}}\email{dalila.failli@unifi.it}

\author[2]{\fnm{Bruno} \sur{Arpino}}\email{bruno.arpino@unipd.it}
\equalcont{These authors contributed equally to this work.}

\author[1]{\fnm{Maria Francesca} \sur{Marino}}\email{mariafrancesca.marino@unifi.it}
\equalcont{These authors contributed equally to this work.}

\affil*[1]{\orgdiv{Dipartimento di Statistica, Informatica, Applicazioni}, \orgname{Università degli Studi di Firenze}, \orgaddress{\street{viale Morgagni 59}, \city{Firenze}, \postcode{50134}, \state{}\country{Italia}}}

\affil[2]{\orgdiv{Dipartimento di Scienze Statistiche}, \orgname{Università degli Studi di Padova}, \orgaddress{\street{Via Battisti 241}, \city{Padova}, \postcode{35121}, \state{}\country{Italia}}}



\abstract{The digital divide is the gap among population sub-groups in accessing and/or using digital technologies. Typically, older people show a lower propensity to have a broadband connection, use the Internet, and adopt new technologies than the younger ones. Motivated by the analysis of the heterogeneity in the use of digital technologies, we build a bipartite network concerning the presence of various digital skills in individuals from three different European countries: Bulgaria, Finland, and Italy. Bipartite networks provide a useful structure for representing relationships between two disjoint sets of nodes, formally called sending and receiving nodes. The goal is to perform a clustering of individuals (sending nodes) based on their digital skills (receiving nodes) for each country. In this regard, we employ a Mixture of Latent Trait Analyzers (MLTA) accounting for concomitant variables, which allows us to (i) cluster individuals according to their profile; (ii) analyze how socio-economic and demographic characteristics, as well as intergenerational ties, influence individual digitalization. Results show that the type of digitalization substantially depends on age, income and level of education, while the presence of children in the household seems to play an important role in the digitalization process in Italy and Finland only.}




\maketitle

\section{Introduction}\label{sec1}
The digital divide is commonly known as the gap among population sub-groups in accessing and/or using digital technologies. This phenomenon has narrowed in recent decades (Hunsaker and Hargittai 2018; Barbosa and Neves 2019), especially as a result of the social restrictions of COVID-19, which increased the use of video calls and instant messaging for a large percentage of older people (Arpino et al. 2020). However, this gap still remains substantial with older people showing a lower propensity to have a broadband connection, use the Internet, and adopt new technologies (Friemel
2016; Di Pietro 2021; Elena-Bucea et al. 2021). Reducing the digital divide could have multiple benefits, such as improving connectivity among people, cutting costs in both business sectors and daily life, and providing access to online public and private services.\\Motivated by the analysis of the heterogeneity in the use of digital technologies, we focus on three European countries: Bulgaria, Finland, and Italy. These are selected on the basis of their degree of digitalization, as measured by the Digital Economy and Society Index (DESI) of the European Commission (2022). This is based on 2021 data and tracks EU Member States' progress in digitalization, showing that, although they made progress in their digitalization efforts during the COVID-19 pandemic, significant gaps in digital skills and digital transformation still need to be addressed. This is particularly relevant for Bulgaria and Italy, which show a low and medium DESI index, equal to about 40\% and 50\%, respectively. In contrast, Finland shows the highest DESI index (about 70\%).\\To analyze the digital divide in these countries, data can be examined by means of bipartite networks. These are particular types of networks that represent relationships between two disjoint sets of nodes, called sending and receiving nodes, respectively. In the present framework, these terms are used to distinguish the two non-overlapping sets of individuals (sending nodes) and digital skills (receiving nodes) {characterizing the bipartite network}.\\{Bipartite networks are increasingly employed in a variety of fields, and are also widely applied in the social and behavioral research, where the main objective is to describe or predict the behaviour of units within the network through their connections with another set of units. An example of application entails the analysis of the participation of society women to a set of social events (Davis, Gardner and Gardner 1941), or the behaviours adopted by
patients to avoid infection during a pandemic (Failli et al.
2022).\\An important aspect of bipartite network analysis often concerns the identification of homogeneous clusters of nodes. This is the case of standard Stochastic Block Models (SBMs; Holland et al. 1983; Snijders and Nowicki 1997; Nowicki et al. 2001; Bartolucci et al. 2018; Marino and Pandolfi 2022), which allow to identify clusters of nodes sharing similar relational profiles. Similarly, the latent position cluster model (Handcock
et al. 2007) identifies communities of strongly connected nodes, , while Aitkin et al. (2014, 2017) exploit a latent class model to identify latent groups of sending nodes in bipartite networks. With the same clustering purposes, Gollini and Murphy (2014)
and Gollini (2020) developed the Mixtures of Latent Trait Analyzers (MLTA) model, which allows to group sending nodes into homogeneous classes, as well as to model the heterogeneity of the connections between sending and receiving nodes. In detail, the MLTA allows to identify latent groups of nodes sharing unobserved characteristics, as in the latent class framework (Bartholomew et al. 2011), and to capture the latent variability of network connections within each group, as in the latent trait framework  (Bartholomew et al. 2011). The combination of these features makes it possible to overcome the problem related to the local independence assumption upon which the latent class model is based (Bartholomew et al. 2011) and which may lead to the identification of an excessive number of groups and, thus, to a difficult interpretation of the results.\\However, the original specification of this model does not account for the effect that nodal attributes, i.e. observed characteristics of sending nodes, may have on clustering formation. Following the standard terminology of latent class models, these can be also referred to as concomitant variables  (Gr\"un and Leisch 2008). To this aim, Failli et al. (2022) extend the MLTA model to explicitly account for the effect that concomitant variables may have on the clustering structure. In this respect, a multinomial logit specification for the prior probabilities of the latent classes is considered.\\In this paper, we focus on this latter specification to analyze the digital divide in Europe. In detail, we focus on information on digital social contacts in work and family life from the European Social Survey (ESS round 10 2020) for Finland, Italy and Bulgaria, with the aim of i) clustering individuals on the basis of their level of digital skills; ii) understanding how individuals' socio-economic and demographic characteristics, as well as intergenerational ties, influence the individual profile of digitalization.\\The inclusion of such covariates within the model is supported by results of previous studies. More specifically, research shows that support from social network can influence the use of digital technologies  (Korupp and Szydlik 2005; Correa et al. 2015; Friemel 2016; Dolnicar et al. 2018; K\"onig et al. 2018; Seguì et al. 2019; Huxhold et al. 2020). Furthermore, some previous studies show that intergenerational learning may facilitate the achievement of digital and social inclusion of older people  (Ivan et al. 2017; Arpino et al. 2020; Azevedo and Ponte 2020; Leek and Rojek 2021; Arpino et al. 2022). Besides social factors, age, income and level of education also play an important role in the individual's level of digitalization (see e.g. Friemel 2016).\\To achieve the above-mentioned objectives, ESS data from Finland, Italy and Bulgaria can be naturally organized in the form of a bipartite network, where sending nodes correspond to individuals and receiving nodes to the digital skills they have. To the best of our knowledge, this methodology has never been applied to the analysis of digital skills.\\ The paper is organized as follows. In Section 2, we describe the ESS data for each country considered in the analysis. In Section 3, we briefly present the MLTA model with concomitant variables. Section 4 presents the application of such model to the ESS data for each country. Section 5 contains concluding remarks and further extensions of the analysis. 

\section{Data}
\label{sec:mot}
In this work, we focus on data from round 10 of the ESS, which covers 21 countries and includes a module based on digital social contacts in work and family life, upon which our analysis is developed. In detail, the ESS is an academic transnational survey whose main objective is to analyze the evolution of attitudes in Europe. The quality of the survey is ensured by random sampling, high response rates, and strict translation protocols (ESS round 10 2020).\\We consider data from three European countries characterized by different levels of digitalization: Finland, Italy, and Bulgaria. As already pointed out in the previous section, these are chosen on the basis of their degree of digitalization measured by the DESI index (European Commission 2022), which is higher for Finland, medium for Italy, and lower for Bulgaria.\\Our goal is to identify groups of individuals with similar digital skills, while also understanding how socio-economic and demographic characteristics, as well as intergenerational ties, influence the individual digitalization profile. To these aims, we build a bipartite network related to the presence of various digital skills for each country. In this framework, sending nodes represent individuals, while receiving nodes represent the digital skills they have. These latter are obtained from information on the frequency of use or familiarity of individuals with respect to specific digital skills:
\begin{enumerate}
    \item Internet use;
    \item Preference settings use;
    \item Advanced search use;
    \item PDFs use;
    \item Video calls with children aged 12 or over, parents, manager, or colleagues;
    \item Messages (via text, email or messaging apps) with children aged 12 or over, parents, manager, or colleagues;
    \item Sharing of online posts about politics in the last 12 months.
\end{enumerate}
For each of the 7 items, the ESS questionnaire ask respondents to report the frequency of use (``never", ``occasionally", ``a few times a week", ``most days", ``every day") or the degree of familiarity (``not at all", ``not very", ``somewhat", ``very", ``completely") with the digital tools or operations. Figures 1, 2, and 3 in the Supplementary Materials show the distributions of these variables for Finland, Italy and Bulgaria, respectively. From here, we can see that in each country most of the individuals considered in the analysis use the \textit{Internet} every day, although the percentage of such people is higher in Finland than in the other two countries. More advanced digital skills, such as the use of \textit{preference settings} and \textit{advanced search}, are less used. Again, the country with the highest use of these skills is Finland. A similar reasoning applies to the use of \textit{PDFs} too. As far as \textit{video calls} and \textit{messages} are concerned, Italy and Bulgaria are the countries where the percentage of respondents who do not use this type of digital communication with colleagues or managers is the highest, while the situation seems to be quite similar as regards video calls and messages with children under 12 and parents. Finally, Finland is the country where sharing \textit{online political posts} is slightly more widespread.
For the purpose of constructing the bipartite network we assume that a tie between pairs of nodes exists if an individual reports to frequently use or to be somewhat familiar with a certain digital technology or skill. Therefore, each of the 7 digital technology variables is transformed into a dummy variables.\\In Table 1, we report the percentage of individuals presenting a tie with different skills, across countries. From this table, it is evident that, as expected, \textit{internet use} is much more widespread in Finland than in Italy and Bulgaria. With regard to \textit{preference settings}, 83\% of Finnish respondents are familiar with them, whereas this percentage drops to 60\% and 50\% in Italy and Bulgaria, respectively. The percentage of individuals who are familiar with \textit{advanced search} is slightly lower than previous percentages, but still higher in Finland than in the other two countries. A similar situation can also be observed for the percentage of respondents who are familiar with \textit{PDFs}. In this case, 55\% of individuals in Bulgaria are not familiar with them. As far as \textit{video calls} are concerned, they are slightly more common in Finland than in the other two nations; Bulgaria is the one presenting the lowest percentage also in this case. The same path is also observed for the skill \textit{messages} and the sharing of \textit{online political posts}. Overall, these results reflect the difference in the digitalization level of the three countries, as highlighted by the DESI index. Similar results are mentioned in Abendroth et al. (2023), where it is evident that digital contacts at work are not yet widespread, especially in Bulgaria, and that countries such as Italy and Finland are characterized by more frequent communication with children than parents, as it is also evident from Figures 1, 2, and 3. 
\begin{table}[ht!]
\caption{Distribution of digital skills variables.}
    \centering
    \begin{tabular}{l c c c}
     \noalign{\smallskip}\hline\noalign{\smallskip}
    & \multicolumn{3}{c}{Country} \\
   Variable  & Finland & Italy & Bulgaria\\
 \noalign{\smallskip}\hline\noalign{\smallskip}
 {Internet use} & 0.91 & 0.75 & 0.69 \\
 \noalign{\smallskip}\hline\noalign{\smallskip}
{Preference settings} & 0.83 & 0.60 & 0.50\\
\noalign{\smallskip}\hline\noalign{\smallskip}
{Advanced search} & 0.79 & 0.60 & 0.52\\
\noalign{\smallskip}\hline\noalign{\smallskip}
{PDFs} & 0.75 & 0.57 & 0.45\\
\noalign{\smallskip}\hline\noalign{\smallskip}
{Video calls} & 0.40 & 0.33 & 0.32\\
\noalign{\smallskip}\hline\noalign{\smallskip}
{Messages} & 0.80 & 0.56 & 0.36\\
\noalign{\smallskip}\hline\noalign{\smallskip}
{Online posts} & 0.20 & 0.17 & 0.10\\
\noalign{\smallskip}\hline\noalign{\smallskip}
    \end{tabular}
    \label{tab:digskill}
\end{table}
\noindent
Considering only individuals with no missing records, we obtain a bipartite network made of $1446$ individuals (i.e., sending nodes) for Finland, $1465$ individuals for Italy, and $2082$ individuals for Bulgaria.\\The data set also contains information on socio-demographic attributes, such as age, education, and health, as well as variables measuring the existence of intergenerational ties, such as the presence of children in the household. The distribution of these variables is shown in Figures 4, 5, and 6 of the Supplementary Materials.
These variables are then transformed into a suitable number of dummy variables or into variables with a smaller number of categories with respect to the original data set ones, for the ease of interpretation, as listed in Table 2. This table shows that the sample considered in the analysis includes both young and older people, as it is evident from the distribution of the variable ``age" in each country. Furthermore, the majority of respondents is healthy, not hampered in daily activities, and is born in the country of residence. However, it can be noted that the percentage of healthy respondents is slightly lower in Bulgaria than in the other two countries, while the percentage of people hampered in daily activities is higher in Finland. Finally, Italy is the country with the highest percentage of respondents who are not born in the country of residence among those considered in the analysis. In addition, it can be seen that respondents are equally distributed with regard to their gender, while the educational level is higher in Finland and lower in Italy. This also holds for the partner's educational level. As far as income is concerned, this is higher in Finland, while there is no substantial difference between Italy and Bulgaria. Furthermore, we note that most individuals in the sample have or have had children in the household in Finland and Bulgaria, while only the 58\% of respondents in Italy have or have had children in the household. As regards the employment status, the percentage of employed people is higher in Finland and lower in Italy, while the percentage of retired people is higher in Bulgaria and lower in Italy. 
\begin{table}[ht!]
\caption{Distribution of demographic and socio-economic variables.}
    \centering
    \begin{tabular}{l l c c c}
      \noalign{\smallskip}\hline\noalign{\smallskip}
    & & \multicolumn{3}{c}{Country}\\
        \noalign{\smallskip}\hline\noalign{\smallskip}
    Variable & Category & Finland & Italy & Bulgaria\\
    \cmidrule(lr){3-5}
\multirow{5}{*}{Age} & $< 30$ & 0.13 & 0.15 & 0.11\\
& $30\vdash50$ & 0.29 & 0.27 & 0.30\\
& $50\vdash65$ & 0.25 & 0.29 & 0.25\\
& $65\vdash75$ & 0.19 & 0.16 & 0.21\\
& $\geq75$ & 0.14 & 0.12 & 0.14\\    \noalign{\smallskip}\hline\noalign{\smallskip}
     \multirow{3}{*}{Health} & Bad & 0.04 & 0.04 & 0.11\\
     & Fair & 0.28 & 0.25 & 0.27\\
      & Good & 0.68 & 0.71 & 0.62\\
       \noalign{\smallskip}\hline\noalign{\smallskip}
    {Daily activity hampered} & Yes & 0.37 & 0.15 & 0.20\\
       \noalign{\smallskip}\hline\noalign{\smallskip}
      {Born in country} & Yes & 0.96 & 0.94 & 0.99\\
    \noalign{\smallskip}\hline\noalign{\smallskip}
  \multirow{2}{*}{Gender} & Male & 0.49 & 0.49 & 0.46\\ 
  & Female & 0.51 & 0.51 & 0.54\\
    \noalign{\smallskip}\hline\noalign{\smallskip}
    \multirow{3}{*}{Education} & Low & 0.14 & 0.38 & 0.17\\
  & Medium & 0.34 & 0.40 & 0.48\\
    & High & 0.52 & 0.22 & 0.35\\ 
    \noalign{\smallskip}\hline\noalign{\smallskip}
    \multirow{3}{*}{Partner's education} & No partner & 0.38 & 0.44 & 0.45\\
    & Low & 0.07 & 0.25 & 0.07\\
   & Medium & 0.22 & 0.21 & 0.29\\
    & High & 0.34 & 0.11 & 0.19\\ 
    \noalign{\smallskip}\hline\noalign{\smallskip}
    \multirow{3}{*}{Household’s net income} & Low & 0.21 & 0.34 & 0.35\\
   & Medium & 0.31 & 0.36 & 0.33\\
    & High & 0.49 & 0.30 & 0.32\\ 
   \noalign{\smallskip}\hline\noalign{\smallskip}
    \multirow{2}{*}{Presence of children} & {Children} & 0.68 & 0.58 & 0.76\\
    & No children & 0.32 & 0.42 & 0.24\\ 
   \noalign{\smallskip}\hline\noalign{\smallskip}
   \multirow{3}{*}{Main activity} & Employed & 0.51 & 0.48 & 0.49\\ 
     &  Unemployed & 0.14 & 0.31 & 0.15\\
    & Retired & 0.34 & 0.21 & 0.36\\
   \noalign{\smallskip}\hline\noalign{\smallskip}
    \end{tabular}
    \label{descrittive2}
\end{table}

\section{Model-based clustering for bipartite networks}
\label{sec:2}
As already pointed out, the research objective is not only to group individuals according to their level of digital skills, but also to understand how socio-economic characteristics and intergenerational ties influence such a level. To these aims, a proper tool of analysis is given by the MLTA model with concomitat variables proposed by Failli et al. (2022). In the following, we briefly illustrate model assumptions and parameter estimation.

\subsection{Model assumptions}
\label{subsec:2}
Let $\mathcal{N} = \{n_1, n_2, \dots ,n_N\}$ denote the set of sending nodes (the individuals) and $\mathcal{R} = \{r_1, r_2, \dots ,r_R\}$ the set of receiving nodes (the digital skills), with $i=1, \dots, N, k=1, \dots, R$, respectively. The relationship structure of a bipartite network can be formally described by a random {incidence matrix} $\boldsymbol{Y} = \{Y_{ik}\}$, whose generic element is given by
\[ 
Y_{ik}=\Bigg\{\begin{array}{@{}l@{}}
    1 \quad \mbox{ if sending node } n_i \mbox{ is connected with receiving node } r_k,\\
    0 \quad \mbox{ otherwise.}
  \end{array}
\]
The standard MLTA model introduced by Gollini and Murphy (2014) and Gollini (2020) assumes that every sending node belongs to an unobserved group identified by the latent random variable $\boldsymbol{z}_i=(z_{i1}, \dots, z_{iG})^{\prime} \stackrel{iid}\sim \mbox{Multinomial}(1,(\eta_{1}, \dots, \eta_{G})^{\prime})$, whose generic element is given by
\[ 
z_{ig}=\Bigg\{\begin{array}{@{}l@{}}
    1 \quad \mbox{ if sending node } n_i \mbox{ belongs to group } g,\\
    0 \quad \mbox{ otherwise.}
  \end{array}
\]
The parameter $\eta_g$ denotes the prior component probability for a randomly selected sending node to belong to group $g$, with $g = 1, \dots ,G$, under the constraints that $\sum_{g=1}^G\eta_g=1$ and $\eta_g\geq0$, $\forall \; g=1, \dots, G$.\\To account for effects of nodal attributes on nodes' clustering, Failli et al. (2022) relaxed the homogeneity assumption of prior component probabilities $\eta_g,\; g = 1, \dots, G,$ and considered the following multinomial latent class specification (Agresti 2002):
\begin{equation*}
\label{nodaleq}
    {\eta{{(\boldsymbol{x}_i; \boldsymbol{\beta}_g)=\text{Pr}(z_{ig}=1|\boldsymbol{x}_i;\boldsymbol{\beta}_g)}}}=\frac{\exp\{\boldsymbol{x}^{\prime}_i\boldsymbol{\beta}_g\}}{1+\sum_{g=2}^G\exp\{\boldsymbol{x}^{\prime}_i\boldsymbol{\beta}_{g}\}}, \quad g=2, \dots, G,
\end{equation*}
where the ${\beta}_{jg}$ coefficient measures the effect of the $j$-th nodal attribute included in the vector $\boldsymbol{x}_i$, with $j=1,\dots,J$, on the logit of belonging to the $g$-th group with respect to the first (the reference one).\\Furthermore, the model assumes the existence of a continuous $D$-dimensional latent trait $\boldsymbol{u}_i$, which is distributed according to a {$D$-variate} Gaussian density, with zero mean vector and identity covariance matrix, i.e., $\boldsymbol{u}_i \sim \mathcal{N}_D(\boldsymbol{0}, \boldsymbol{I})$. Latent traits represent theoretical constructs or hidden features that are assumed to influence the connections between sending and receiving nodes. Therefore, the size $D$ of the latent trait denotes the number of underlying traits that the model attempts to capture within the set of receiving nodes considered in the analysis.\\Given the previous assumptions, the model postulates that the conditional distribution of the vector $\boldsymbol{Y}_i = (Y_{i1}, \dots, Y_{iR})^{\prime}$, given that node $n_i$ belongs to the $g$-th group, is specified by a latent trait model with parameters $b_{gk}$ and $\boldsymbol{w}_{gk}$, with $g = 1, \dots, G$, and $k = 1, \dots, R$. In this sense, conditional on $\boldsymbol{z}_i$ and $\boldsymbol{u}_i$, variables contained in the $\boldsymbol{Y}_i$ vector are assumed to be independent Bernoulli random variables with parameters $\pi_{gk}(\boldsymbol{u}_i)$, $k = 1, \dots, R$. These are modelled via the following logistic function:
\begin{equation}
\label{pi}
    \pi_{gk}(\boldsymbol{u}_{i})=\text{Pr}(Y_{ik}=1\mid \boldsymbol{u}_{i}, z_{ig}=1)=\frac{1}{1+\exp[-(b_{gk}+\boldsymbol{w}_{gk}^{\prime}\boldsymbol{u}_{i})]}, \quad 0\leq \pi_{gk}(\boldsymbol{u}_{i})\leq 1.
\end{equation}
Here, $b_{gk}$ is the model intercept and represents the \textit{attractiveness} of receiving node $r_k$ for sending nodes belonging to the $g$-th group, with $g=1,\dots,G$. The larger the value of this parameter, the higher the attractiveness of the $k$-th receiving node for those sending nodes in the $g$-th component; on the contrary, the low its value, the lower the attractiveness and, thus, the lower the propensity of sending nodes in the $g$-th component to be connected to the $k$-th receiving node. On the other side, $\boldsymbol{w}_{gk}$ are the slopes associated with the latent trait $\boldsymbol{u}_i$ and are meant to capture the \textit{influence} of the latent trait on the probability of a connection between receiving node $r_k$ and the sending nodes belonging to the $g$-th group. Statistically significant $\boldsymbol{w}_{gk}$ parameters indicate association between receiving nodes, as well as the presence of heterogeneity within groups with respect to the baseline level dictated by $b_{gk}$. In detail, a high value for this parameter denotes greater differences in the probability of a connection between sending and receiving nodes. Therefore, the sign of this coefficient mainly denotes the nature of the influence (positive or negative) of the latent trait on the probability of a connection between sending and receiving nodes.\\A simpler and more parsimonious version of the model is obtained by assuming constant $\boldsymbol{w}_{gk}$ parameters across groups ($\boldsymbol{w}_{gk} = \boldsymbol{w}_k \; \forall \; g=1,\dots,G$), meaning that the latent trait has the same effect in all groups.

\subsection{Model estimation}
Regarding the estimation of model parameters, a standard solution when dealing with latent variables relies on the use of an EM algorithm (Dempster et al. 1977). However, in this framework, it requires the evaluation of a {multidimensional integral which does not admit a closed-form expression, as it is evident from the following model log-likelihood function 
\[
\ell(\boldsymbol{\theta}) = \sum_{i=1}^N \log \Bigg(\sum_{g=1}^G \eta(\boldsymbol{x}_i;\boldsymbol{\beta}_g)\int_{\mathbb{R}^D}f(y_{ik}\mid \boldsymbol{u}_i, z_{ig}=1)f(\boldsymbol{u}_i)d\boldsymbol{u}_i \Bigg).
\]
Here, $\boldsymbol{\theta}$ represents the vector of all free model parameters and $f(y_{ik}\mid \boldsymbol{u}_i, z_{ig}=1)=\pi_{gk}(\boldsymbol{u}_i)^{y_{ik}}(1-\pi_{gk}(\boldsymbol{u}_i))^{(1-y_{ik})}$. To solve the issue,  Gollini and Murphy (2014) propose to combine the EM algorithm with a variational approximation of the multidimensional integral (Jaakkola and Jordan 1997; Tipping 1998). The goal is to approximate the component densities, $f(y_{ik} \mid \boldsymbol{u}_i, z_{ig}=1)$, determining the log-likelihood function with a lower bound depending quadratically in the exponential on the latent trait $\boldsymbol{u}_i$. This is then conjugate to the Gaussian density postulated for $\boldsymbol{u}_i$, and allows the multidimensional integral to admit a closed-form solution. The resulting estimation procedure corresponds to a double EM algorithm with a {variational approximation} of the likelihood function, alternating the following steps until convergence:
\begin{enumerate}
    \item E-step: compute the expected value of the complete data log-likelihood function, conditional on the observed data and the current value of parameter estimates. In the $(t+1)$-th iteration, this is equivalent to compute the posterior probabilities of the latent variables $z_{ig}$.
    \item M-step: the multinomial logit coefficients $\boldsymbol{\beta}_g$ are estimated via a Newton-Raphson algorithm by maximizing the likelihood of a multinomial logit model, with weights provided by $\hat{z}_{ig}^{(t+1)}$ derived at the E-step. Prior probabilities ${\eta{{(\boldsymbol{x}_i; \boldsymbol{\beta}_g)}}}$ are updated accordingly.
    \item The integral in the complete-data log-likelihood is approximated by using the variational approach. To this aim, a second EM algorithm, nested within the first, is required:
\begin{itemize}
    \item E-step: compute the approximate posterior distribution of the continuous latent trait $\boldsymbol{u}_i$, given the observation $\boldsymbol{y}_i$, the posterior probability $\hat{z}_{ig}^{(t+1)}$, and the variational parameters $\hat{\boldsymbol{\xi}}_{ig}^{(t)}$.
    \item M-step: the variational parameters $\boldsymbol{\xi}_{ig}$ are estimated by maximizing the expected value of the complete data log-likelihood function with respect to each $\xi_{igk} \in \boldsymbol{\xi}_{ig}$, $k = 1, \dots, R$. Updates for $\boldsymbol{w}_{gk}$ and $b_{gk}$ are obtained similarly.
    \end{itemize}
\end{enumerate}
For further details on the estimation method we refer  to Gollini and Murphy (2014) and Failli et al. (2022).\\As typically happens with latent variables, an initialization strategy based on multiple starting points may help the estimation algorithm avoid being trapped in local maxima solutions. Once model parameters are estimated, each sending node is assigned to one of the $G$ groups on the basis of the estimated posterior probabilities $\hat{z}_{ig}$, $g=1, \dots, G$, via a Maximum a Posteriori (MAP) rule.

\subsection{Standard errors}
\label{subsec:4}
To provide and estimate of model parameters' standard errors, Failli et al. (2022) suggest the use of a {non-parametric bootstrap} approach (Efron 1979). Given an incidence matrix $\boldsymbol{Y}$, this method consists in drawing with repetition $S$ samples having the same size of $\boldsymbol{Y}$. In detail, for each bootstrap replicate, $N$ rows of the incidence matrix are drawn with repetition, so that each sending node can appear several times and an MLTA model is estimated for fixed $G$ and $D$.\\Let $\hat{\boldsymbol{\theta}}_{(s)}$ denote the vector of estimates obtained from the $s$-th bootstrap sample. By paying attention to the label switching issue, bootstrap standard errors correspond to the square root of the diagonal elements of the following matrix:
\[
    \mbox{V}(\hat{\boldsymbol{\theta}})=\frac{1}{S}\sum_{s=1}^S \big(\hat{\boldsymbol{\theta}}_{(s)}-\hat{\boldsymbol{\theta}}_{(.)}\big)\big(\hat{\boldsymbol{\theta}}_{(s)}-\hat{\boldsymbol{\theta}}_{(.)}\big)^{\prime},
\]
with $\hat{\boldsymbol{\theta}}_{(.)}$ being the empirical mean vector $\hat{\boldsymbol{\theta}}_{(.)}=\frac{1}{S}\sum_{s=1}^S \hat{\boldsymbol{\theta}}_{(s)}.$

\section{The digital divide network}
\label{sec:3}
In this section, we apply the MLTA model with concomitant variables to the ESS data for Finland, Italy and Bulgaria. Results for each country are presented in the following sections. Note that, to avoid the estimation algorithm from being trapped in local maxima, a multi-start strategy based on 10 random starts is considered. As regards the cocomitant variables, we decided to focus on the categorical variables described in Section 2. The model for each country is estimated for a number of groups, $G$, ranging from 1 to 4 and a latent trait size, $D$, ranging from 0 to 3. Both the unconstrained (i.e., the model with different $\boldsymbol{w}_{gk}$ across groups) and the constrained model (i.e., the model with constant $\boldsymbol{w}_k$ across groups) are considered. The model corresponding to the smallest value of the Bayesian Information Criterion (BIC; Schwarz 1978) is selected as the optimal one.

\subsection{Results from Finland}\label{fin_mod}
Tables 3 and 4 show the BIC values for Finland, for the unconstrained and constrained model, respectively. The optimal specification is the one corresponding to $G=2$ groups, a one-dimensional latent trait ($D=1$), and different $\boldsymbol{w}_{gk}$ across groups (Table 3).
\begin{table}[ht!]
\caption{BIC for unconstrained MLTA (different $\boldsymbol{w}_{gk}$ across groups) and for varying numbers of groups $G$ and latent trait dimensions $D$ (Finland).}
    {\begin{tabular}{{lrrrr}}
    \toprule
     & \textit{G}=1 & \textit{G}=2 & \textit{G}=3 & \textit{G}=4 \\
    \midrule
    \textit{D}=0 &  8744.80 &  16103.64 & 16283.59 & 16299.93 \\
    \textit{D}=1  & 8744.79 & \textbf{7931.08} & 8009.96 & 8100.53\\
    \textit{D}=2  & 8715.03   & 7988.37 & 8093.67 & 8329.72\\
    \textit{D}=3 & 8848.27  & 8078.93 & 8214.62 & 8433.85\\
    \bottomrule
    \end{tabular}}
        \label{bic1}
\end{table}
\begin{table}[ht!]
\caption{BIC for constrained MLTA (constant $\boldsymbol{w}_k$ across groups) and for varying numbers of groups $G$ and latent trait dimensions $D$ (Finland).}
    {\begin{tabular}{{lrrrr}}
    \toprule
     & \textit{G}=1 & \textit{G}=2 & \textit{G}=3 & \textit{G}=4 \\
     \midrule
     \textit{D}=0 &  8744.80 &  16103.64 & 16283.59 & 16299.93 \\
    \textit{D}=1  & 8744.79 & 7967.73 & 8030.17 & 7996.66\\
    \textit{D}=2 &  8715.03   & 8013.47 & 7976.59 & 8039.76\\
    \textit{D}=3 & 8848.27   & 8064.31 & 8110.20 & 8074.98\\
    \bottomrule
    \end{tabular}}
    \label{bic2}
\end{table}\\A relevant aspect of the analysis concerns the interpretation of attractiveness coefficients $b_{gk}$ and the influence parameters $\boldsymbol{w}_{gk}$, $g=1, \dots, G$, $k=1, \dots, R$. Estimates and 95\% confidence intervals for such parameters are shown in Figure 7 and 8, respectively. As outlined in Section 3.1, positive (respectively, negative) values for these parameters indicate a high (respectively, low) attractiveness of the $k$-th digital skill for individuals belonging to the $g$-th group. By looking at Figure 7, we notice that groups can be labeled according to their level of digitalization, which is lower in the first group and higher in the second. In detail, only the attractiveness coefficients related to \textit{Internet use} and \textit{text messaging} are statistically significant and positive in group 1, thus indicating that individuals in this group mainly have these two skills. In contrast, the only negative coefficients in group 2 are those related to the sharing of \textit{online political posts} and to the making of \textit{video calls}, meaning that individuals in this group are disinclined to have these two digital skills. Furthermore, we can see that the most commonly employed digital skill in this group is the \textit{Internet} use. This may be partly due to the fact that \textit{Internet} use is also a precondition for several of the other skills.
\begin{figure}[ht!]
   \begin{minipage}{0.48\textwidth}
     \centering
     \includegraphics[width=1\linewidth]{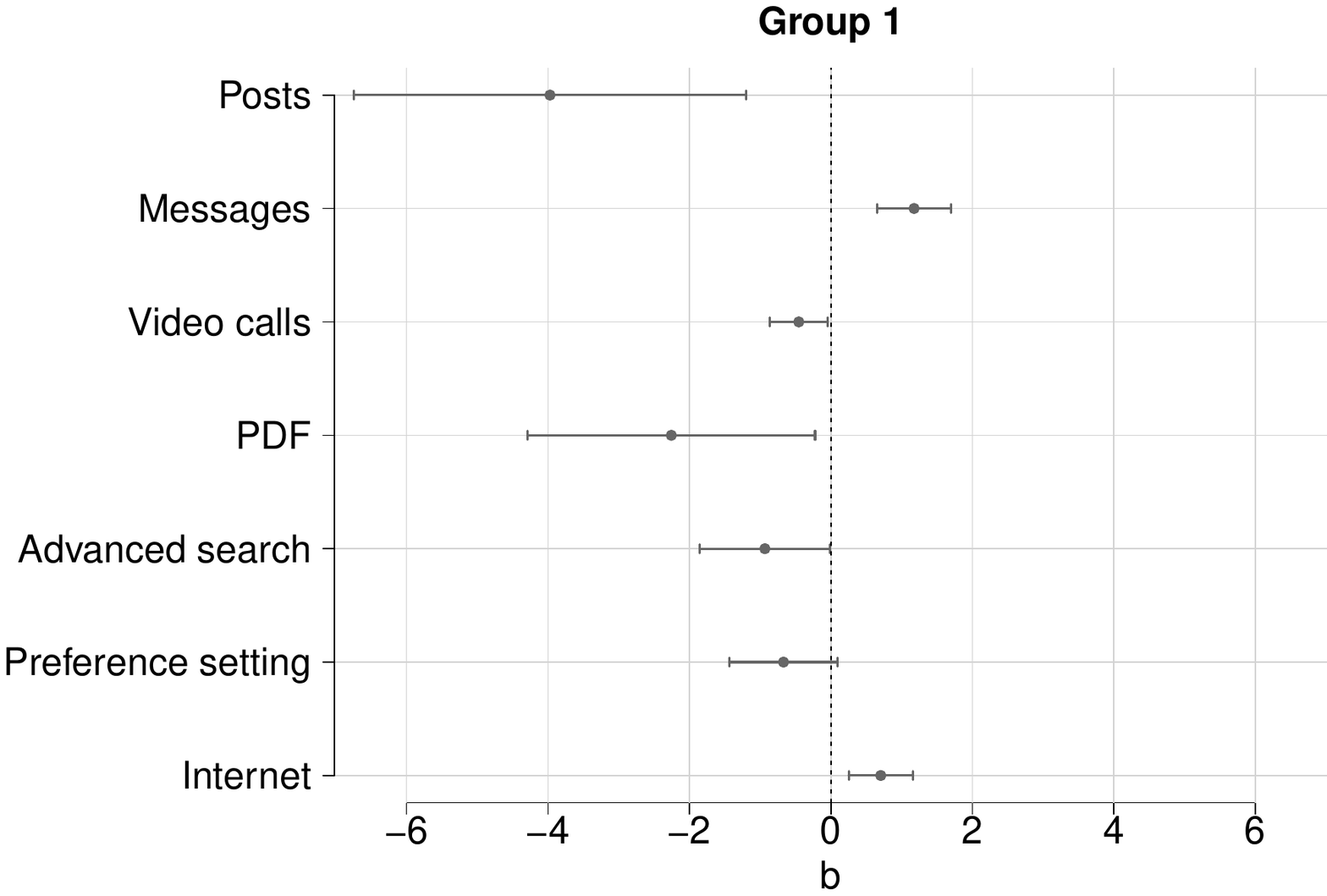}
   \end{minipage}\hfill
   \begin{minipage}{0.48\textwidth}
     \centering
     \includegraphics[width=1\linewidth]{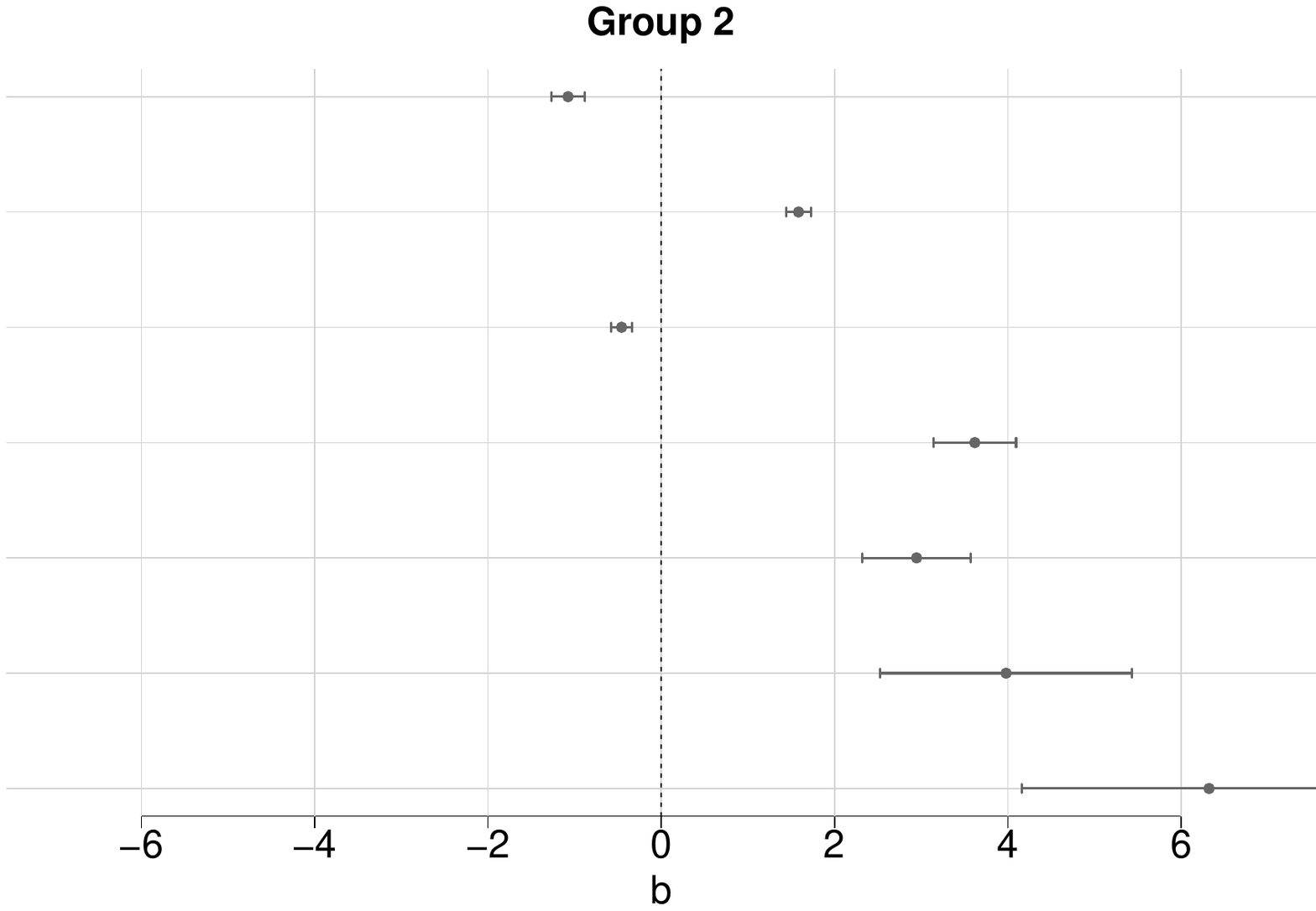}
   \end{minipage}
   \caption{Attractiveness coefficients estimates $\hat{b}_{gk}$ and 95\% bootstrap confidence intervals (Finland).}
\label{fin_b}       
\end{figure}\\A measure of the influence of the latent trait on the probability of having a specific digital skill for individuals belonging to the $g$-th group is given by ${w}_{gk}$ parameters. As highlighted in Section 3.1, a statistically significant effect of these parameters indicates not only association between digital skills, but also the presence of unexplained heterogeneity within groups with respect to the baseline level $b_{gk}$. Figure 8 represents the estimates and the 95\% confidence intervals for each of such coefficients. By looking at them, it is evident that the least digitalized group presents greater variability in the use of digital tools, given the high significance of all the ${w}_{gk}$ coefficients, for $g = 1$, and $k = 1, \dots, R$. Pairing these results with the estimated values of the attractiveness parameters $b_{gk}$, $g = 1$ and $k = 1, \dots, R$, we may conclude that less digitalized individuals show greater variability in digital skills with respect to the more digitalized ones (i.e., individuals in group 2).
\begin{figure}[ht!]
   \begin{minipage}{0.48\textwidth}
     \centering
     \includegraphics[width=1\linewidth]{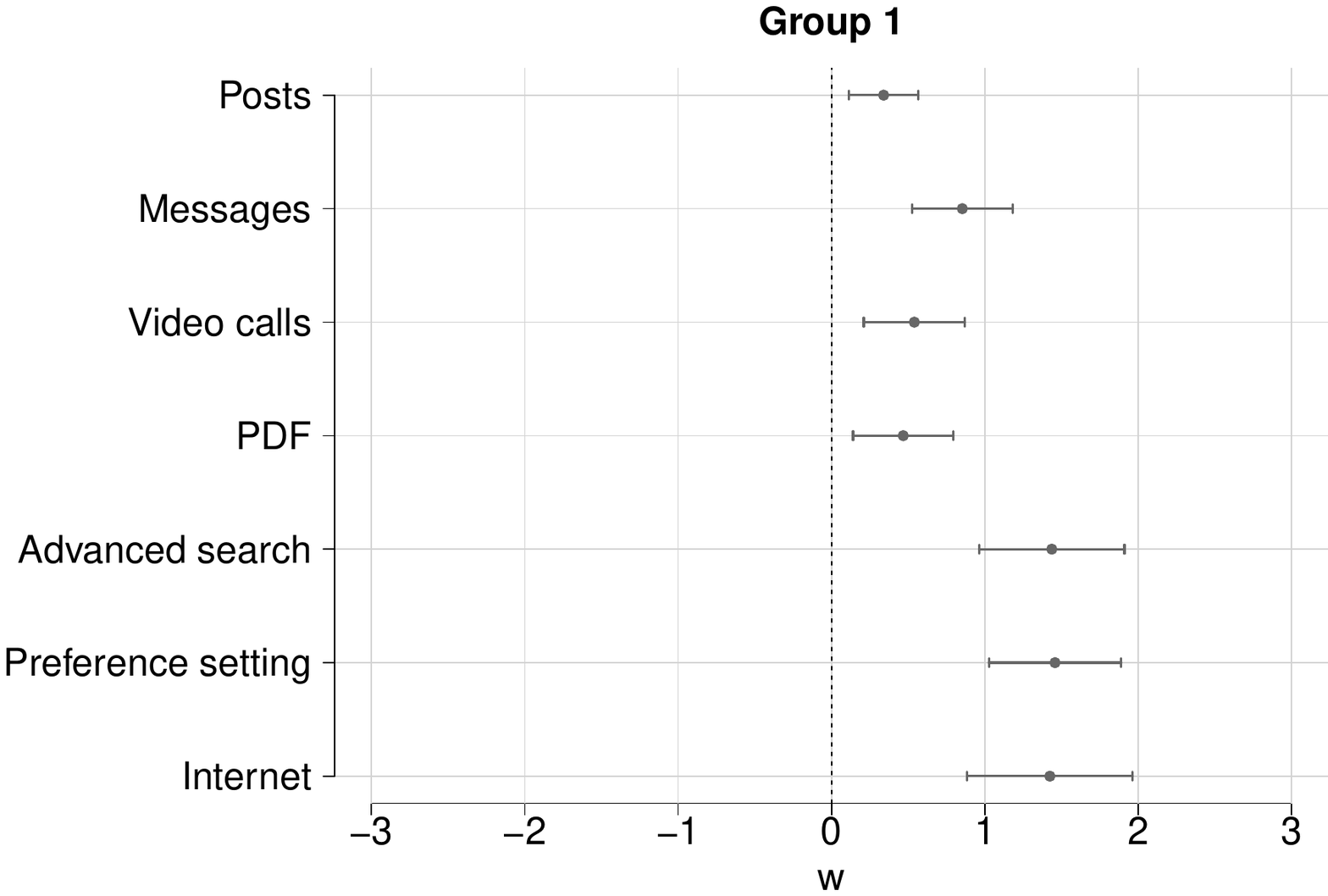}
   \end{minipage}\hfill
   \begin{minipage}{0.48\textwidth}
     \centering
     \includegraphics[width=1\linewidth]{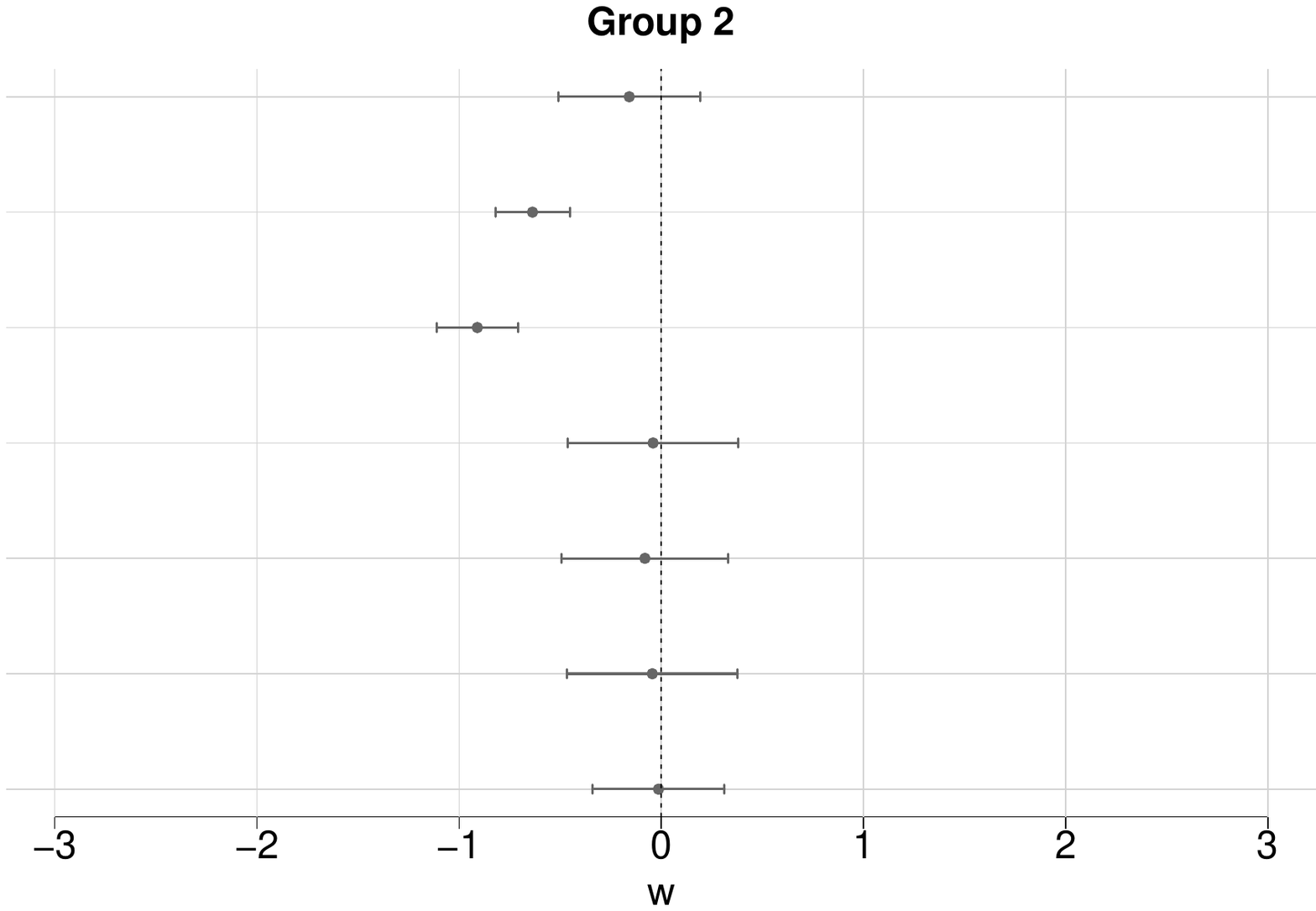}
   \end{minipage}
   \caption{Influence parameters estimates $\hat{w}_{gk}$ and 95\% bootstrap confidence intervals (Finland).}
\label{fin_w}
\end{figure}\\These estimates can be used to compute the predicted probabilities of having each digital skill for individuals belonging to the $g$-th group, by using Equation (1). Figure 9 in the Supplementary Materials shows the predicted probabilities' distributions and the corresponding mean (red line). By looking at this figure, it is evident that, in the first group, the chance of having any of the the digital skills considered in this analysis tends to be lower than in the second group, apart from using \textit{video calls} and \textit{text messages}. The most owned digital skills for individuals belonging to the first group are the most basic ones, i.e., general \textit{internet use} and \textit{text messages}. All other digital skills are rather uncommon for individuals in this group. On the contrary, group 2 is characterized by the use of both basic and more advanced digital technologies, with an average probability of being owned above 0.8, apart from the use of \textit{video calls} and sharing \textit{political posts}. The predicted probability for the latter is still much higher than that observed for the first group. 
\\Regarding the effect of nodal attributes on nodes clustering, Table 5 shows the estimated $\boldsymbol{\beta}_g$ coefficients measuring the impact of individuals' socio-economic and demographic features, as well as intergenerational ties, on the logit of belonging to cluster $g=1, 2$, together with the corresponding 95\% confidence interval. Here, the reference group corresponds to the least digitalized group ($g=1$). By looking at the sign and at the significance of these coefficients, we may conclude that the chance of belonging to the most digitalized group is higher for individuals in good health and who had children in the household, while it is lower for those who are over 65 years of age, for those with a low/medium educational level, for those whose partner has a low level of education, for those without partner (compared to those whose partner has a high level of education), and for people with a low/medium income.
\begin{table}[ht!]
\footnotesize
    \centering
\caption{Estimated $\boldsymbol{\beta}_g$ parameters and 95\% bootstrap CIs (Finland). The reference category is in brackets.}
    \begin{tabular}{l r r}
    \noalign{\smallskip}\hline\noalign{\smallskip}
     Variable & $\hat{\boldsymbol{\beta}}_2$ & 95\% CI\\
         \noalign{\smallskip}\hline\noalign{\smallskip}
      Intercept & 0.43 & (0.06; 0.81)\\
      Age $30\vdash50$ ($<$30) & -10.17 & (-23.63; 3.30) \\
      Age $50\vdash65$ ($<$30) & -13.37 & (-27.02; 0.29) \\
      Age $65\vdash75$ ($<$30) & -13.81 & (-27.42; -0.19) \\
      Age $75+$ ($<$30) & -15.71 & (-29.31; -2.10) \\
      Health fair (bad) & 1.23 & (0.40; 2.06) \\ 
      Health good (bad) & 1.37 & (0.51; 2.23) \\
      Hampered & 0.08 & (-0.29; 0.44) \\
      Born in country & -0.34 & (-1.68; 1.00)  \\
      Male & 0.03 & (-0.28; 0.33) \\
      Education low (high) & -2.36 & (-2.90; -1.82) \\ 
      Education medium (high) & -1.60 & (-1.99; -1.21)\\
      Partner education low (high) & -0.73 & (-1.43; -0.03) \\ 
      Partner education medium (high) & -0.21 & (-0.66; 0.24) \\
      Partner education not applicable (high) & -0.99 & (-1.58; -0.40) \\
      Income low (high) & -1.35 & (-1.97; -0.73) \\
      Income medium (high) & -0.87 & (-1.31; -0.42) \\
      Children & 0.71 & (0.21, 1.20) \\
      Retired (employed) & -0.58 & (-1.28; 0.12) \\
      Unemployed (employed) & -0.06 & (-0.69; 0.58) \\     
      \noalign{\smallskip}\hline\noalign{\smallskip}
    \end{tabular}
    \label{fin_beta}
\end{table}\\Finally, Table 6 shows the predicted probabilities of belonging to each group for some covariates of interest, such as age, health, education and income. This table confirms that the probability of belonging to the most digitally skilled group is higher for young, healthy, and well-educated people, as well as for those with higher incomes. For example, the probability of belonging to the most digitalized group ($G$=2) is very close to zero for individuals over 30 years of age.
\begin{table}[ht!]
    \centering
     \caption{Predicted probabilities of belonging to each group for some covariates of interest (Finland).}
     \setlength{\tabcolsep}{2pt}
     \footnotesize{
    \begin{tabular}{{lcccccccccccccccc}}
    \toprule
         &  \multicolumn{16}{c}{Variables}\\
        &  \multicolumn{5}{c}{Age} &  \multicolumn{3}{c}{Health}   &  \multicolumn{3}{c}{Education}   &  \multicolumn{3}{c}{Income}   &  \multicolumn{2}{c}{Children}\\
        \cmidrule(lr){2-6}
        \cmidrule(lr){7-9}
        \cmidrule(lr){10-12}
        \cmidrule(lr){13-15}
        \cmidrule(lr){16-17}
       Group  & $<30$ & $30\vdash50$ & $50\vdash65$ & $65\vdash75$ & $75+$ & Bad & Fair & Good & Low & Medium & High & Low & Medium & High & Yes & No \\
       \cmidrule(lr){1-1}
       \cmidrule(lr){2-6}
        \cmidrule(lr){7-9}
        \cmidrule(lr){10-12}
        \cmidrule(lr){13-15}
        \cmidrule(lr){16-17}
       $G$=1 & 0.01 & 0.99 & 0.99 & 0.99 & 0.99 & 0.83 & 0.23 & 0.20 & 0.91 & 0.83 & 0.01 & 0.79 & 0.70 & 0.04 & 0.33 & 0.79\\
         $G$=2 & 0.99 & 0.01 & 0.01 & 0.01 & 0.01 & 0.17 & 0.77 & 0.80 & 0.09 & 0.17 & 0.99 & 0.21 & 0.30 & 0.96 & 0.67 & 0.21\\
         \bottomrule
    \end{tabular}}
    \label{p2.fin}
\end{table}

\subsection{Results from Italy}\label{it_mod}
Tables 7 and 8 show the BIC values for Italy, for the unconstrained and constrained model, respectively. The optimal specification is the one corresponding to $G=3$, a one-dimensional latent trait ($D=1$), and varying $\boldsymbol{w}_{gk}$ across groups (Table 7). 
\begin{table}[ht!]
\caption{BIC for unconstrained MLTA (different $\boldsymbol{w}_{gk}$ across groups) and for varying numbers of groups $G$ and latent trait dimensions $D$ (Italy).}
    {\begin{tabular}{{lrrrr}}
    \toprule
     & \textit{G}=1 & \textit{G}=2 & \textit{G}=3 & \textit{G}=4\\
    \midrule
    \textit{D}=0 & 12858.12 & 15933.29 & 16031.67 & 16171.95\\
  \textit{D}=1 & 10189.84 & 9047.27 & \textbf{8932.20} & 8976.85\\
    \textit{D}=2 & 10028.03 & 9126.37 & 9079.72 & 9156.69\\
    \textit{D}=3 & 10151.22  & 9218.27 & 9209.41 & 9310.39\\
    \bottomrule
    \end{tabular}}
        \label{bic1ita}
\end{table}
\begin{table}[ht!]
\caption{BIC for constrained MLTA (constant $\boldsymbol{w}_k$ across groups) and for varying numbers of groups $G$ and latent trait dimensions $D$ (Italy).}
    {\begin{tabular}{{lrrrr}}
    \toprule
     & \textit{G}=1 & \textit{G}=2 & \textit{G}=3 & \textit{G}=4 \\
     \midrule
    \textit{D}=1 & 10189.84 & 9000.04 & 9019.43 & 8936.49 \\
    \textit{D}=2 & 10028.03 & 9045.81 & 8941.76 & 8942.51\\
    \textit{D}=3 & 10151.22 & 9098.66 & 8967.57 & 8991.46\\
    \bottomrule
    \end{tabular}}
    \label{bic2ita}
\end{table}\\As for Finland, estimates and 95\% confidence intervals for attractiveness coefficients $b_{gk}$ and influence parameters $\boldsymbol{w}_{gk}$, $g=1, \dots, G$, $k=1, \dots, R$ are shown in Figures 10 and 11, respectively. By looking at this figure, it is evident that individuals in the first group are less inclined to have each of the digital skills considered in the analysis. Attractiveness coefficients are indeed all negative and not statistically significant. Furthermore, individuals in group 2 mainly use the \textit{Internet}. In contrast, individuals in group 3 adopt all digital tools, but they are less inclined to share \textit{online political posts}. 
\begin{figure}[ht!]
   \begin{minipage}{0.33\textwidth}
     \centering
     \includegraphics[width=1\linewidth]{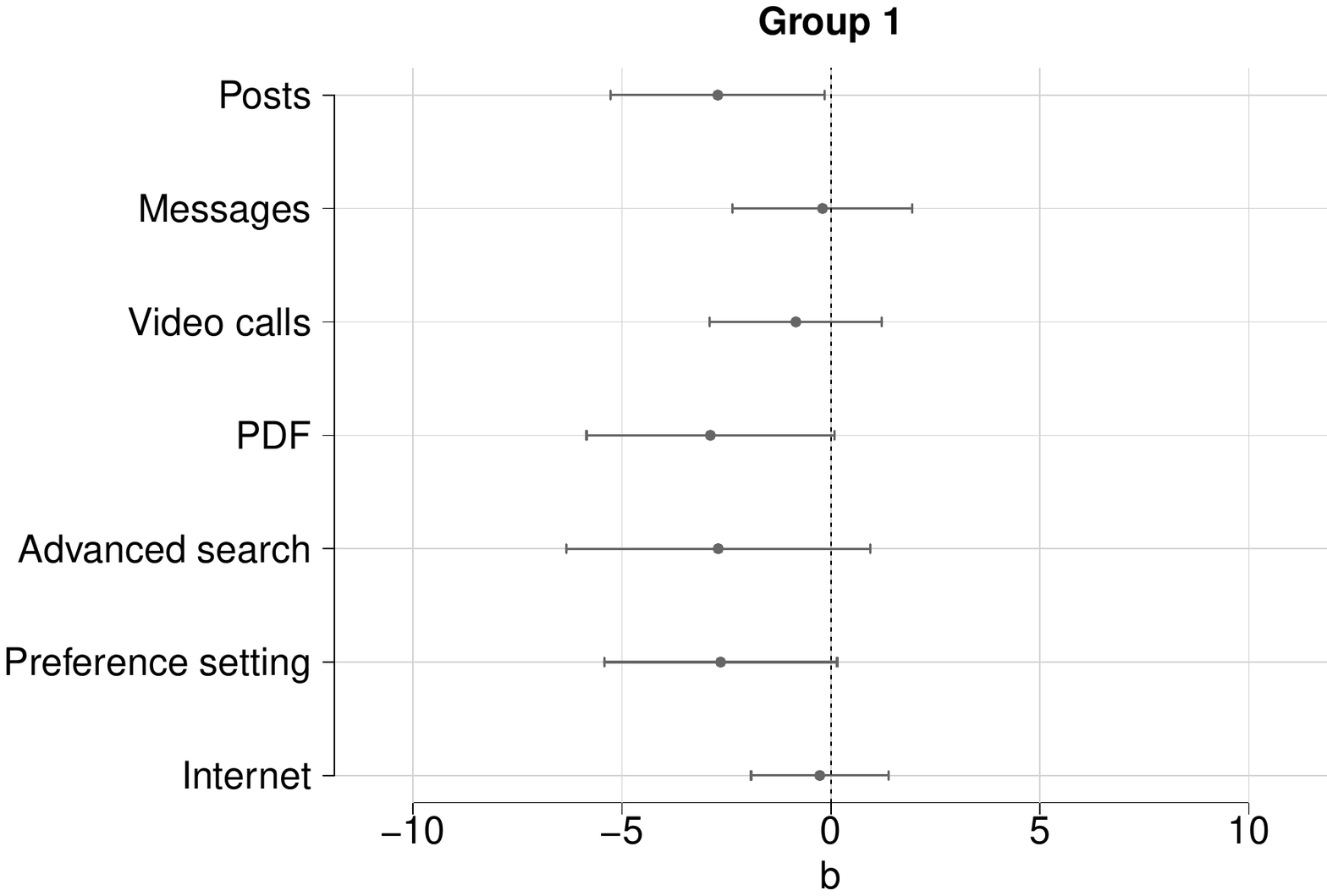}
   \end{minipage}\hfill
   \begin{minipage}{0.33\textwidth}
     \centering
     \includegraphics[width=1\linewidth]{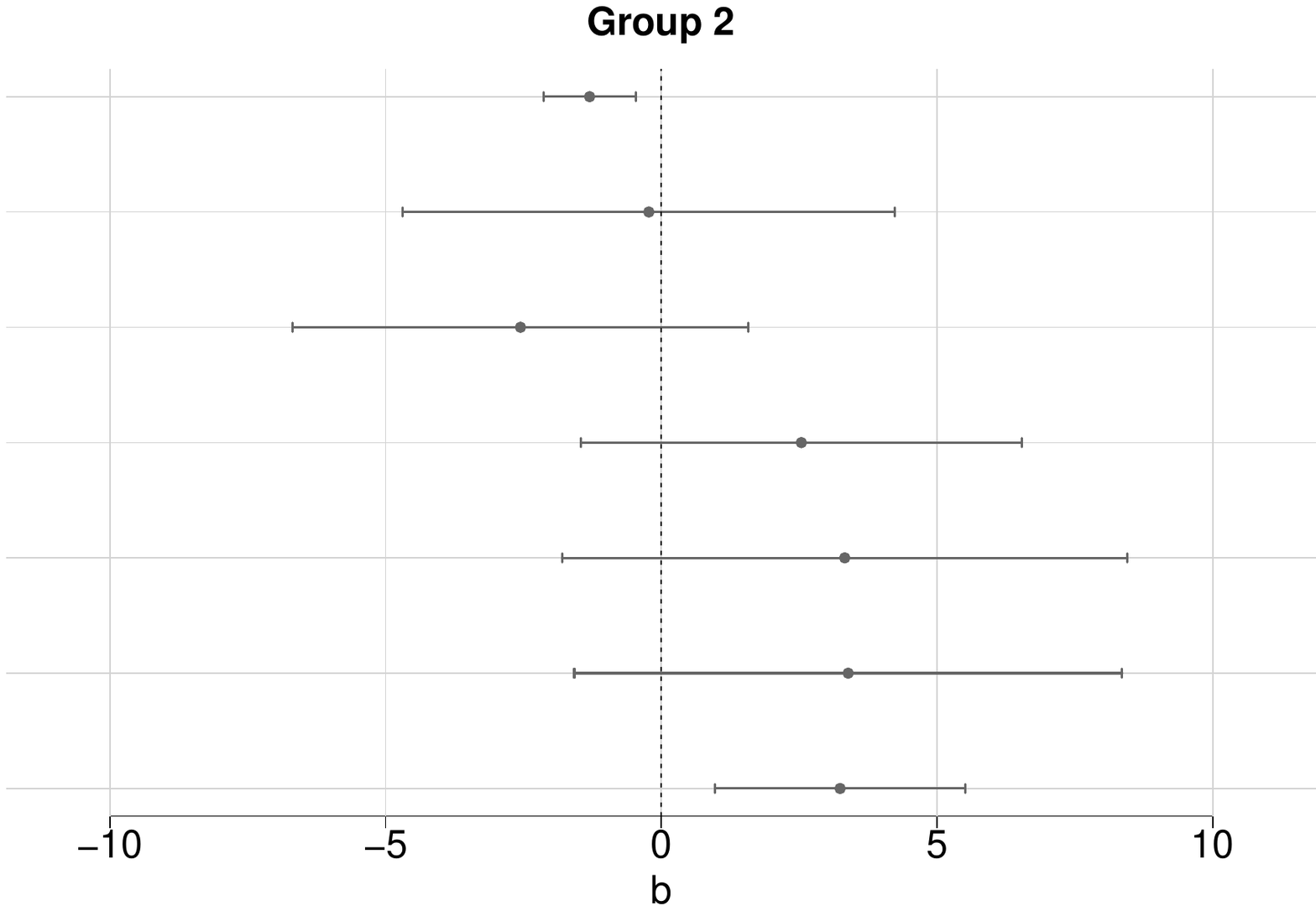}
   \end{minipage}\hfill
   \begin{minipage}{0.33\textwidth}
     \centering
     \includegraphics[width=1\linewidth]{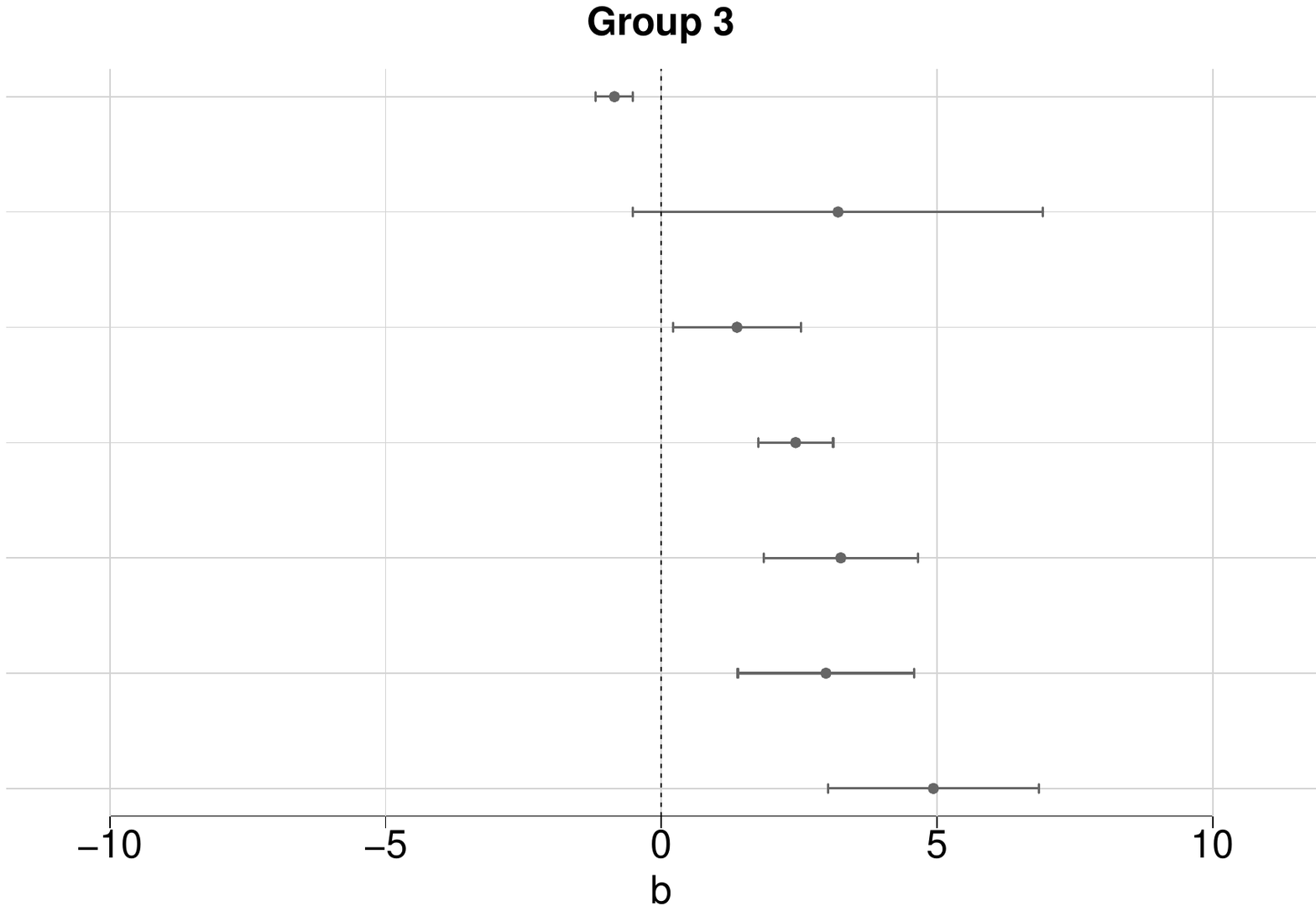}
   \end{minipage}
   \caption{Attractiveness coefficients estimates $\hat{b}_{gk}$ and 95\% bootstrap confidence intervals (Italy).}
\label{b_ita}       
\end{figure}\\Figure 11 represents the estimates and the 95\% confidence intervals for ${w}_{gk}$ parameters. By looking at them, it is evident that the least digitalized group presents greater variability in the use of digital tools with respect to more digitalized groups (2 and 3), these show absence of association between digital skills and lack of heterogeneity with respect to the baseline level. 
\begin{figure}[ht!]
   \begin{minipage}{0.33\textwidth}
     \centering
     \includegraphics[width=1\linewidth]{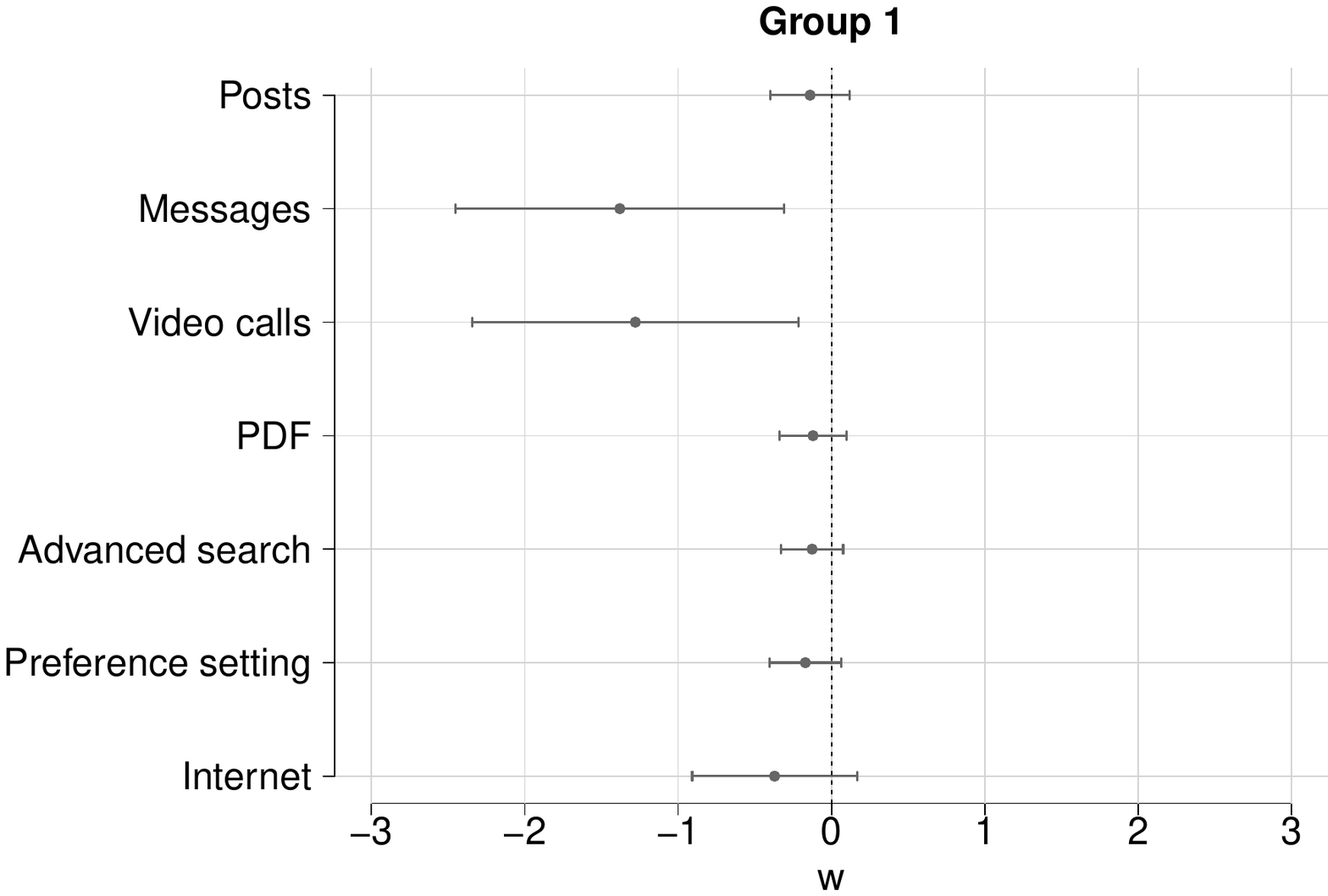}
   \end{minipage}\hfill
   \begin{minipage}{0.33\textwidth}
     \centering
     \includegraphics[width=1\linewidth]{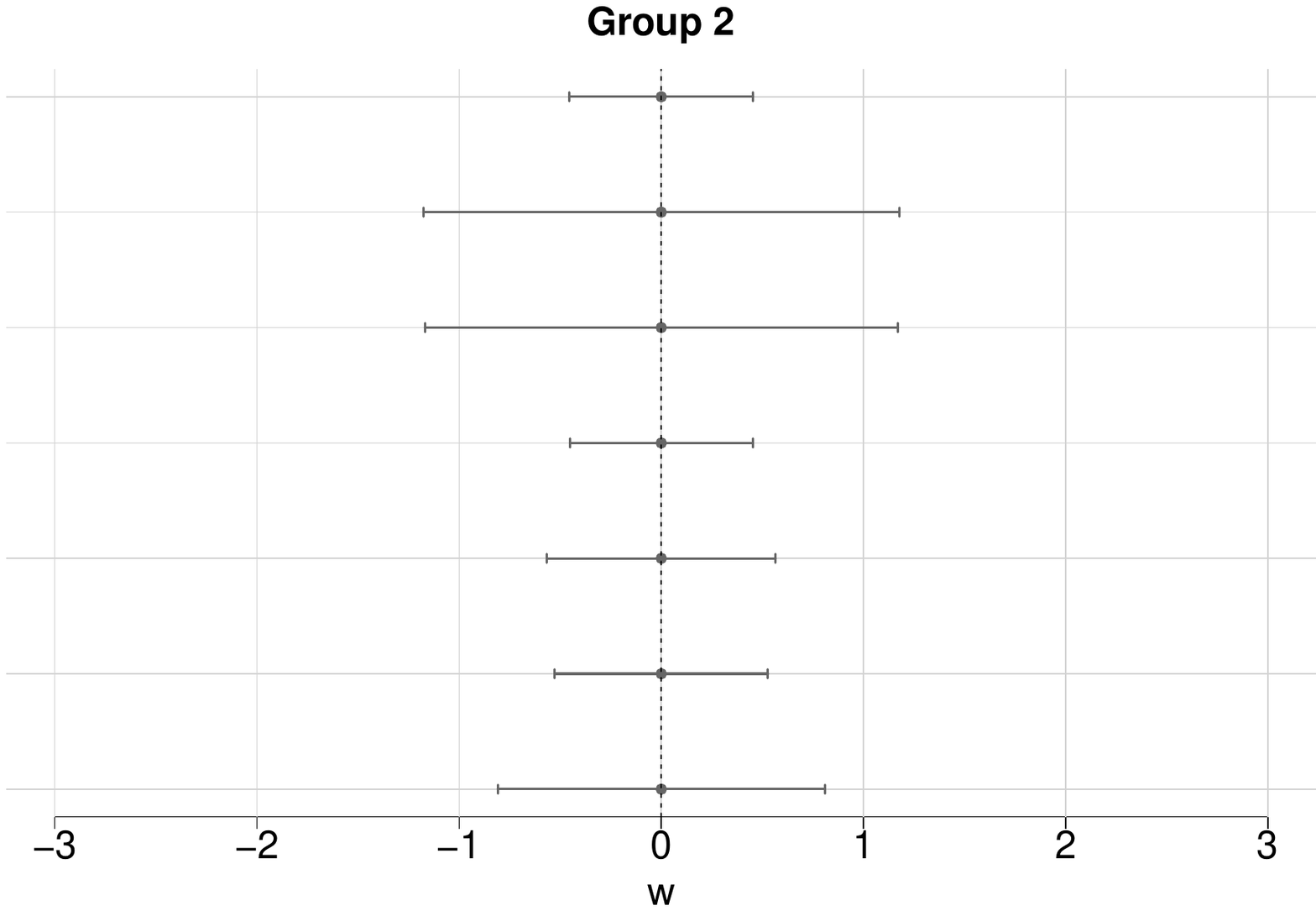}
   \end{minipage}\hfill
   \begin{minipage}{0.33\textwidth}
     \centering
     \includegraphics[width=1\linewidth]{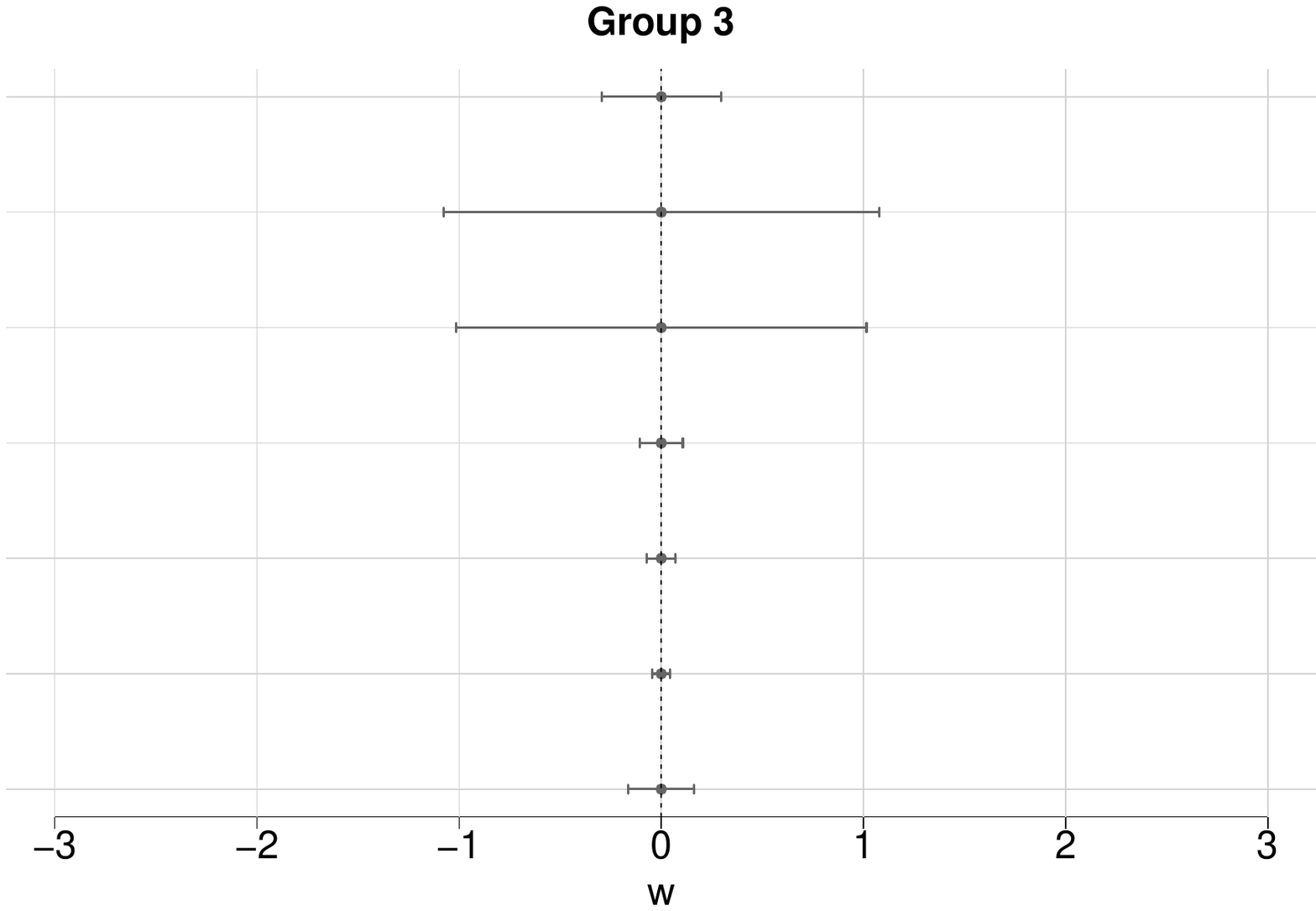}
   \end{minipage}
  \caption{Influence parameters estimates $\hat{w}_{gk}$ and 95\% bootstrap confidence intervals (Italy).}
\label{w_ita}       
\end{figure}\\Figure 12 in the Supplementary Materials shows the distributions of the predicted probabilities of having each digital skill for individuals belonging to the $g$-th group and their corresponding means (red lines). Looking at the figure, it is evident that these probabilities are lower in the first group, although the probability of using \textit{Internet}, \textit{video calls}, and \textit{text messaging} is higher with respect to other skills. On the other hand, in group 2 and 3, all probabilities are rather high and above 0.9. The only exceptions regard the use of \textit{video calls}, sending \textit{text messages} (44\% chance for group 2), and sharing \textit{online political posts} for group 2. Overall, it is evident that groups can be labelled according to the level of digitalization, which is lower in group 1 and higher in group 3.
\\Regarding the effect of nodal attributes on nodes clustering, Table 9 shows the estimated $\boldsymbol{\beta}_g$ coefficients, together with the corresponding 95\% confidence interval. Here, the reference group corresponds to the least digitalized group (i.e., group 1). By looking at these coefficients, we see that the chance of belonging to group 2 with respect to the reference is lower for people over 65 years of age, for those in good health, with a low/medium educational level, with a low income, and for those who are retired or unemployed, while the chance is higher for males. Furthermore, estimates for group 3 show that the chance of belonging to the most digitalized group is lower for those over 30 years of age and with a medium income, while such chances are higher for males and unemployed people. 
\begin{table}[ht!]
\footnotesize
    \centering
\caption{Estimated $\boldsymbol{\beta}_g$ parameters and 95\% bootstrap CIs (Italy). The reference category is in brackets.}
    \begin{tabular}{l r r | r r}
    \noalign{\smallskip}\hline\noalign{\smallskip}
     & $\hat{\boldsymbol{\beta}}_2$ & 95\% CI & $\hat{\boldsymbol{\beta}}_3$ & 95\% CI \\
         \noalign{\smallskip}\hline\noalign{\smallskip}
      Intercept & 6.07 & (2.83; 9.30) & 1.76 & (-8.02; 11.54)  \\
      Age $30\vdash50$ ($<$30) & -1.86 & (-4.57; 0.84) & -2.18 & (-2.96; -1.41) \\
      Age $50\vdash65$ ($<$30) &  -2.27 & (-4.64; 0.11) & -3.44 & (-5.07; -1.80) \\
      Age $65\vdash75$ ($<$30) & -3.56 & (-5.73; -1.39) & -3.19 & (-5.75; -0.63) \\
      Age $75+$ ($<$30) & -3.68 & (-5.98; -1.38) & -5.02 & (-8.72; -1.31) \\
      Health fair (bad) & -1.46 & (-2.19; -0.72) & 1.19 & (-2.39; 4.77)\\ 
      Health good (bad) & -0.96 & (-1.77; -0.15) & 1.81 & (-1.02; 4.63) \\
      Hampered &  -0.48 & (-0.98; 0.03) & -0.22 & (-0.87; 0.42) \\
      Born in country & 0.35 & (-0.24; 0.95) & 0.90 & (-3.40; 5.20) \\
      Male &  0.44 & (0.13; 0.76) & 0.73 & (0.08; 1.38) \\
      Education low (high) & -2.75 & (-3.29; -2.21) & -2.60 & (-5.97; 0.77) \\ 
      Education medium (high) & -1.23 & (-1.65; -0.80) & -1.21 & (-4.51; 2.09) \\
      Partner education low (high) & -0.43 & (-1.26; 0.39) & -0.89 & (-2.26; 0.49) \\ 
      Partner education medium (high) & 0.10 & (-0.53; 0.73) & -0.01 & (-0.69; 0.66) \\
      Partner education not applicable (high) & -0.23 & (-0.95; 0.50) & -0.51 & (-2.27; 1.26) \\
      Income low (high) & -1.07 & (-1.58; -0.56) & -1.34 & (-2.73; 0.06) \\
      Income medium (high) & -0.45 & (-0.95; 0.05) & -1.58 & (-2.58; -0.57) \\
      Children & 0.89 & (-0.06; 1.84) & 2.02 & (-0.03; 4.06) \\
      Retired (employed) & -2.26 & (-2.93; -1.59) & 0.82 & (-1.44; 3.07) \\
      Unemployed (employed) & -1.48 & (-1.94; -1.02) & 0.66 & (0.24; 1.07) \\ 
      \noalign{\smallskip}\hline\noalign{\smallskip}
    \end{tabular}
    \label{beta_ita}
\end{table}
Finally, Table 10 shows the predicted probabilities of belonging to each group for some covariates of interest. This table confirms that the probability of belonging to the most digitally skilled group is higher for young, healthy, and well-educated individuals, as well as for those with higher incomes.  For example, the probability of belonging to the most digitalized group ($G$=3) ranges from 0.09 for individuals between 30 and 50 years of age to 0.01 for individuals aged 75 and over. Furthermore, the probability of belonging to the most digitally qualified group is higher for people with children in the household (0.69) than for those without children (0.15).
\begin{table}[ht!]
    \centering
     \caption{Predicted probabilities of belonging to each group for some covariates of interest (Italy).}
     \setlength{\tabcolsep}{2pt}
    \footnotesize{
    \begin{tabular}{{lcccccccccccccccc}}
    \toprule
         &  \multicolumn{16}{c}{Variables}\\
        &  \multicolumn{5}{c}{Age} &  \multicolumn{3}{c}{Health}   &  \multicolumn{3}{c}{Education}   &  \multicolumn{3}{c}{Income}   &  \multicolumn{2}{c}{Children}\\
        \cmidrule(lr){2-6}
        \cmidrule(lr){7-9}
        \cmidrule(lr){10-12}
        \cmidrule(lr){13-15}
        \cmidrule(lr){16-17}
       Group  & $<30$ & $30\vdash50$ & $50\vdash65$ & $65\vdash75$ & $75+$ & Bad & Fair & Good & Low & Medium & High & Low & Medium & High & Yes & No \\
       \cmidrule(lr){1-1}
       \cmidrule(lr){2-6}
        \cmidrule(lr){7-9}
        \cmidrule(lr){10-12}
        \cmidrule(lr){13-15}
        \cmidrule(lr){16-17}
       $G$=1 & 0.00 & 0.79 & 0.88 & 0.93 & 0.97 & 0.96 & 0.22 & 0.13 & 0.88 & 0.63 & 0.00 & 0.65 & 0.53 & 0.01 & 0.09 & 0.40\\
         $G$=2 & 0.08 & 0.12 & 0.09 & 0.03 & 0.02 & 0.04 & 0.05 & 0.05 & 0.06 & 0.18 & 0.54 & 0.22 & 0.33 & 0.20 & 0.22 & 0.45\\
         $G$=3 & 0.92 & 0.09 & 0.03 & 0.04 & 0.01 & 0.00 & 0.73 & 0.81 & 0.07 & 0.19 & 0.46 & 0.13 & 0.14 & 0.79 & 0.69 & 0.15\\
         \bottomrule
    \end{tabular}}
    \label{p2.ita}
\end{table}

\subsection{Results from Bulgaria}\label{bulg_mod}
Tables 11 and 12 show the BIC values for Bulgaria, for unconstrained and constrained model, respectively. By looking at these tables, we note that the optimal specification is the one corresponding to $G=3$ groups, a one-dimensional latent trait ($D=1$), and constant $\boldsymbol{w}_{k}$ across groups (Table 12). 
\begin{table}[ht!]
\caption{BIC for unconstrained MLTA (different $\boldsymbol{w}_{gk}$ across groups) and for varying numbers of groups $G$ and latent trait dimensions $D$ (Bulgaria).}
    {\begin{tabular}{{lrrrr}}
    \toprule
     & \textit{G}=1 & \textit{G}=2 & \textit{G}=3 & \textit{G}=4 \\
    \midrule
    \textit{D}=0 & 17972.87 & 22270.94 & 22292.32 & 22081.27 \\
  \textit{D}=1 & 13744.22 & 12077.67 & 11770.58 & 11743.42  \\
    \textit{D}=2 & 13473.02 & 12174.46 & 11914.03 & 11737.45 \\
    \textit{D}=3 & 13616.39 & 12294.95 & 12025.79 & 11919.99 \\
    \bottomrule
    \end{tabular}}
        \label{bic1bulg}
\end{table}
\begin{table}[ht!]
\caption{BIC for constrained MLTA (constant $\boldsymbol{w}_k$ across groups) and for varying numbers of groups $G$ and latent trait dimensions $D$ (Bulgaria).}
    {\begin{tabular}{{lrrrr}}
    \toprule
     & \textit{G}=1 & \textit{G}=2 & \textit{G}=3 & \textit{G}=4 \\
     \midrule
    \textit{D}=1 & 13744.22 & 12040.64 & \textbf{11663.9} & 11729.08 \\
    \textit{D}=2 &  13473.02 & 12091.95 & 11730.15 & 11698.32 \\
    \textit{D}=3 & 13616.39 & 12178.31 & 11809.06 & 11813.13 \\
    \bottomrule
    \end{tabular}}
    \label{bic2bulg}
\end{table}\\Estimates and 95\% confidence intervals for attractiveness parameters $b_{gk}$ and influence parameters $\boldsymbol{w}_{k}$ are shown in Figure 13 and 14, respectively. By looking at these estimates, we notice that, as for Finland and Italy, groups can be labeled according to the digitalization level of their members, which is lower in group 1 and higher in group 3. Indeed, the estimated coefficients in group 1 are all lower than zero or not statistically significant, suggesting that individuals in this group are not very digitally literate. A similar consideration, but to a lesser extent, also applies to group 2, where the \textit{Internet use} coefficient is not statistically significant, but its magnitude is relevant. Finally, in group 3, coefficients for \textit{Internet}, \textit{preference setting}, \textit{advanced search} and \textit{PDFs} use are all positive and significant, meaning that individuals in this group are slightly more digitally skilled than the others. 
\begin{figure}[ht!]
   \begin{minipage}{0.33\textwidth}
     \centering
     \includegraphics[width=1\linewidth]{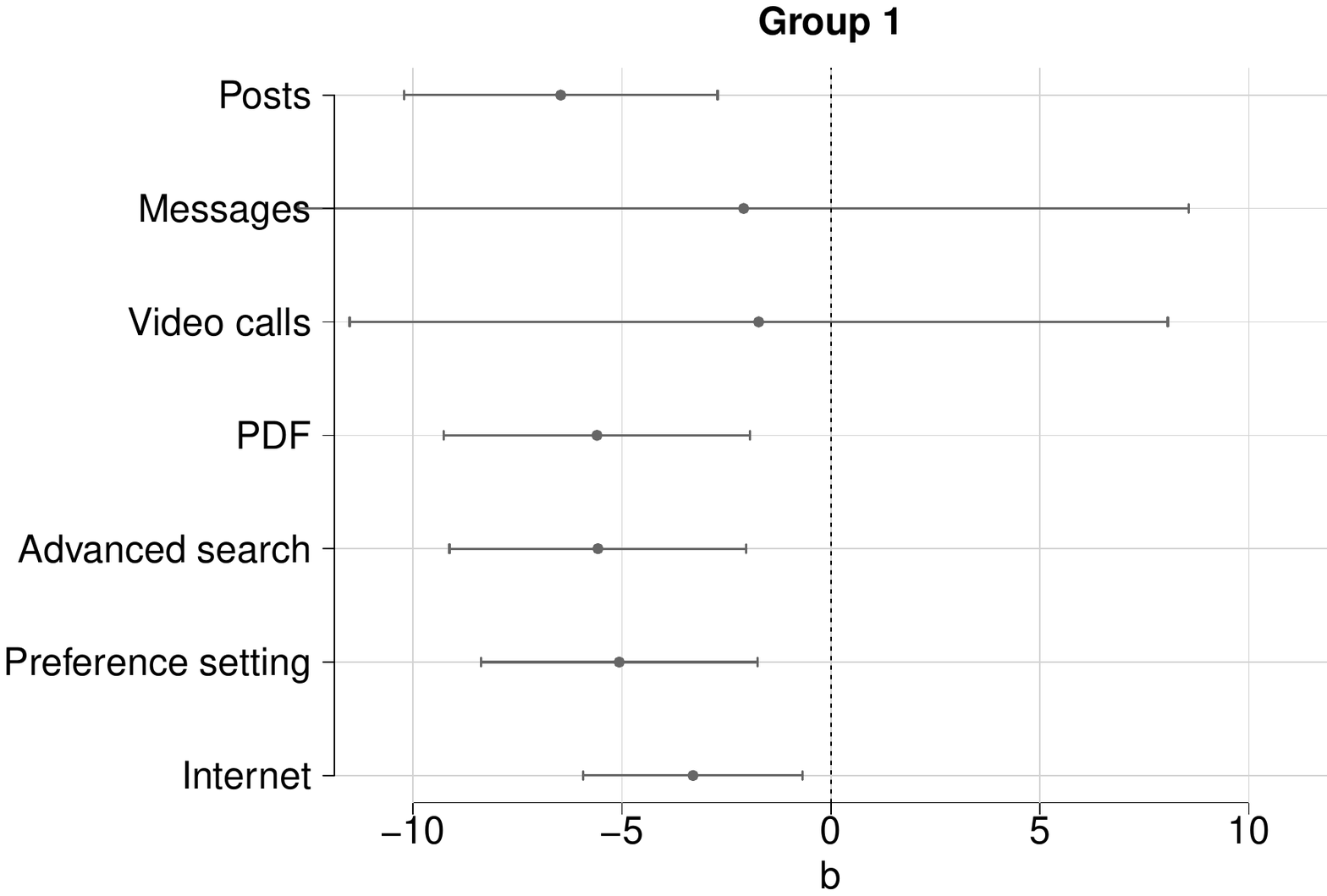}
   \end{minipage}\hfill
   \begin{minipage}{0.33\textwidth}
     \centering
     \includegraphics[width=1\linewidth]{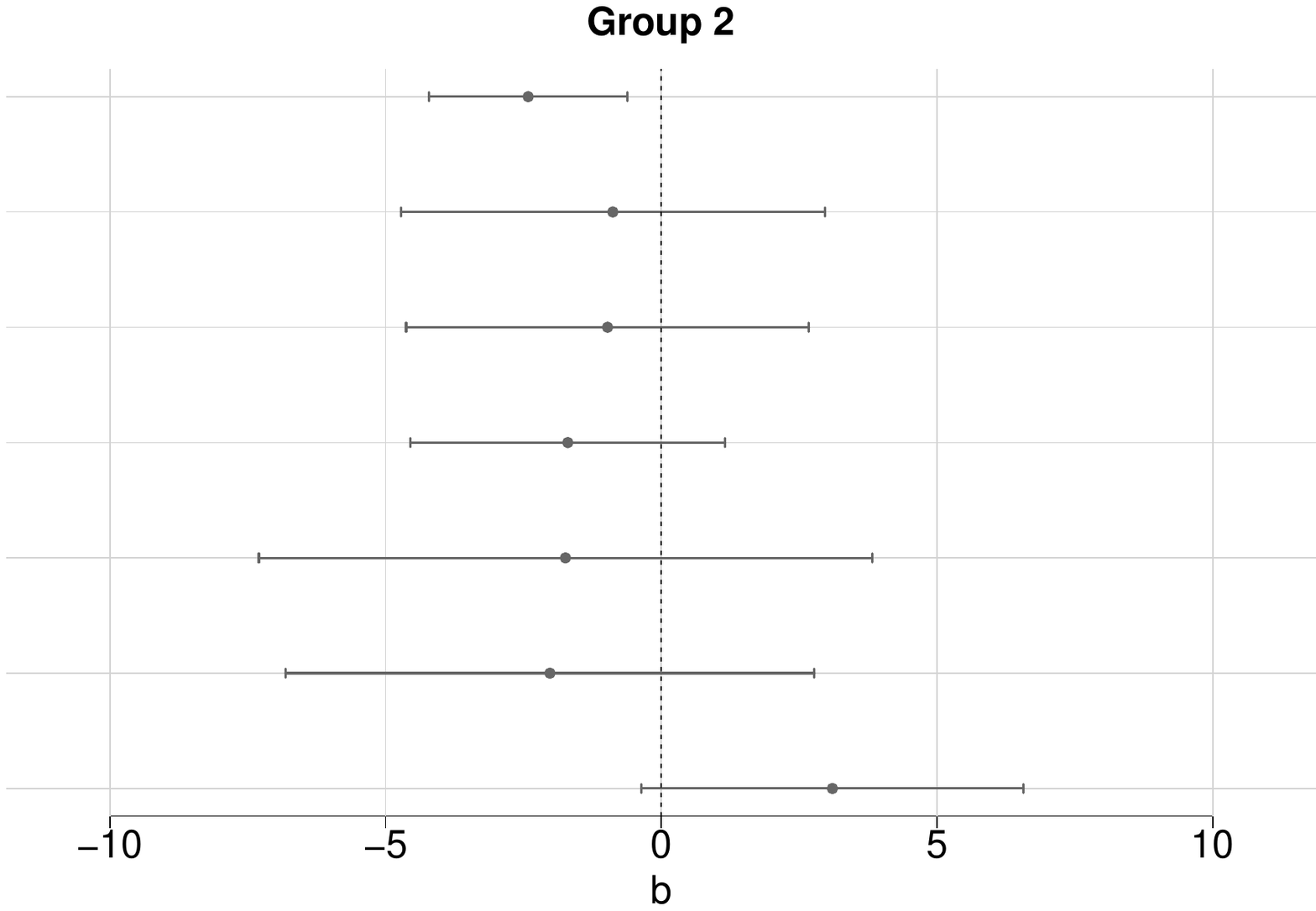}
   \end{minipage}\hfill
   \begin{minipage}{0.33\textwidth}
     \centering
     \includegraphics[width=1\linewidth]{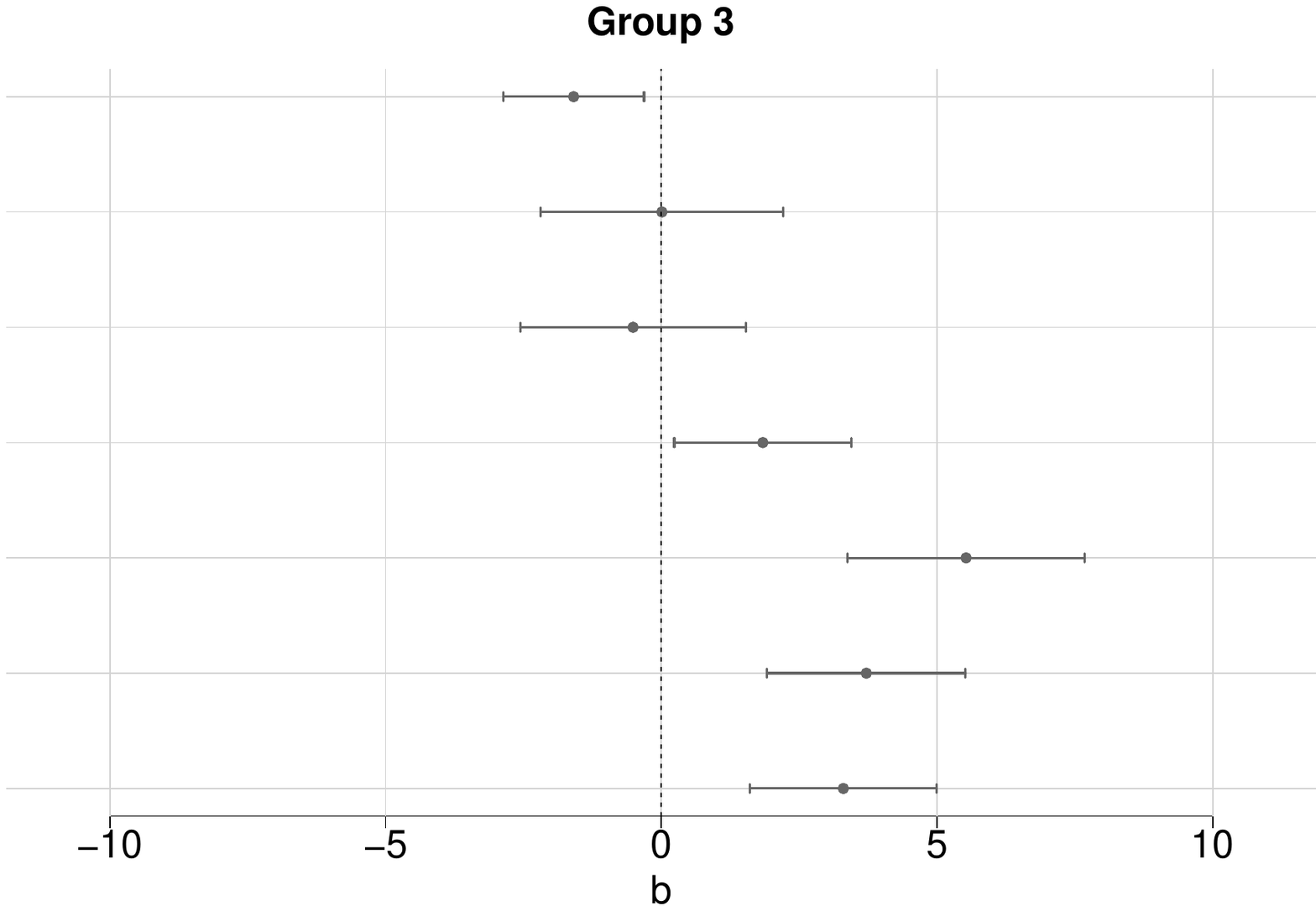}
   \end{minipage}
   \caption{Attractiveness coefficients estimates $\hat{b}_{gk}$ and 95\% bootstrap confidence intervals (Bulgaria).}
\label{bulg_b}       
\end{figure}
\\Figure 14 represents the estimates and the 95\% confidence intervals for ${w}_{k}$ parameters. By looking at them, it is evident that individuals show greater variability in the use of \textit{video calls} and \textit{messages}, while all the other digital skills are characterized by lack of heterogeneity with respect to the baseline level $b_{gk}$.
\begin{figure}[ht!]
     \centering
     \includegraphics[width=0.45\linewidth]{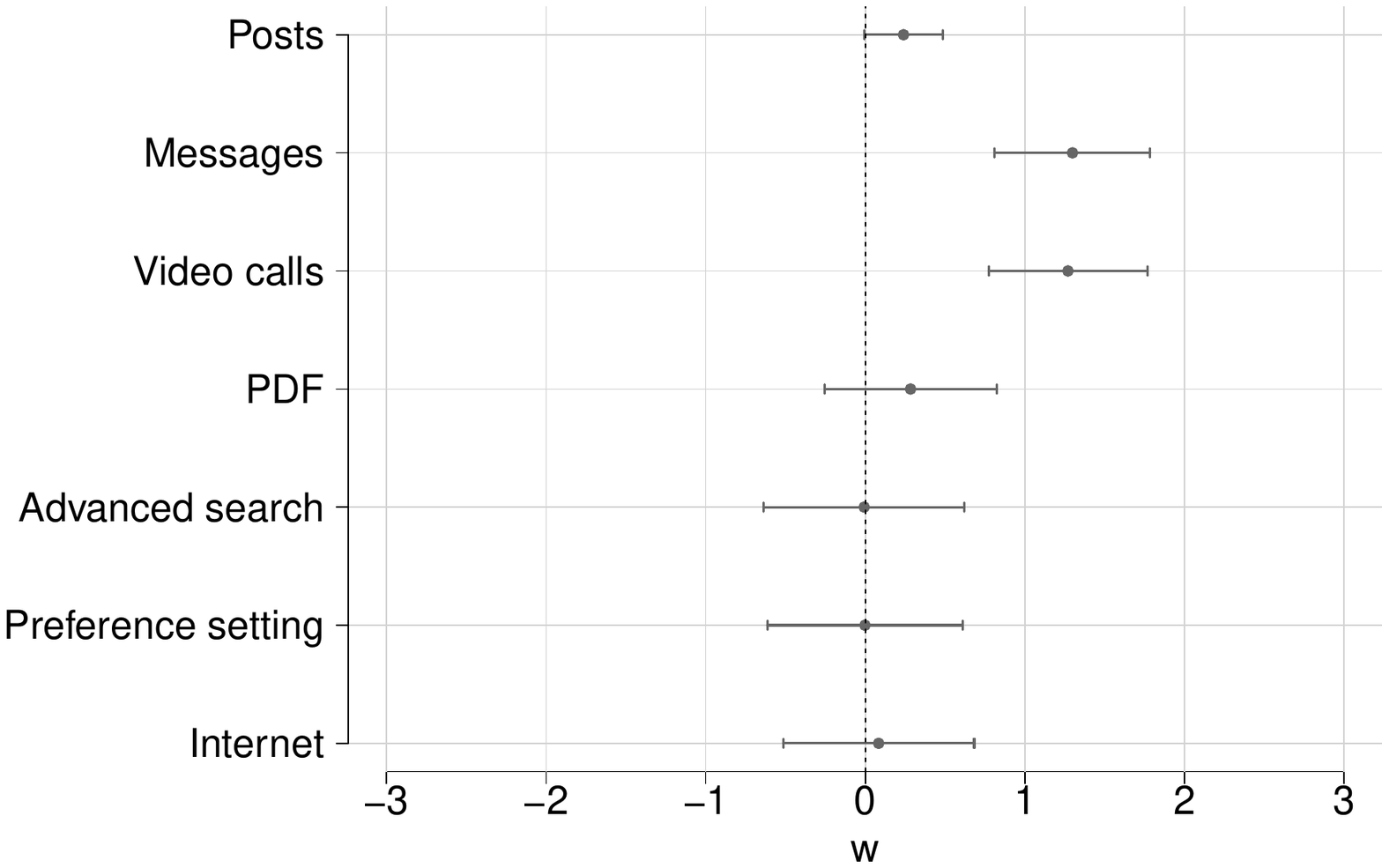}
   \caption{Influence parameters estimates $\hat{w}_{gk}$ and 95\% bootstrap confidence intervals (Bulgaria).}
\label{w_bulg}
\end{figure}\\Figure 15 of the Supplementary Materials shows the distributions of the predicted probabilities of having each digital skill for individuals belonging to the $g$-th group and their corresponding means (red lines). Looking at the figure, it is evident that the first group shows very low probabilities of having the different digital skills considered in this analysis, also compared to Finland and Italy. On the other hand, in group 2, such probabilities are slightly higher, especially with regard to basic skills such as \textit{text messaging}. As for general \textit{Internet use}, the mean predicted probability in group 2 is very high and comparable to that in group 3. Finally, in the third group, all probabilities are very high, while the use of \textit{video calls}, \textit{messages}, and the sharing of \textit{online political posts} are less frequent, albeit the observed probabilities are higher than those observed for the former groups. Therefore, also for this country, groups can be labelled according to the level of digitalization, which is lower in group 1 and higher in group 3.
\\Regarding the effect of nodal attributes on nodes clustering, Table 13 shows the estimated $\boldsymbol{\beta}_g$ coefficients, together with the corresponding 95\% confidence interval. Here, the reference group corresponds to the least digitalized one ($G$=1). By looking at the sign and at the significance of these coefficients, we may conclude that the chance of belonging to the medium digitalized group (i.e., group 2) is lower for those who are aged 50 or over, while it is higher for those in good health. Similarly, the chance of belonging to the most digitalized group (i.e., group 3) is lower for those who are aged 50 or over. The estimates for other covariate show high levels of uncertainty with wide 95\% confidence intervals. However, also in Bulgaria low and medium educational levels are associated with lower chances of belonging to the most digitalized group compared to people with higher education. 
\begin{table}[ht!]
\footnotesize
    \centering
\caption{Estimated $\boldsymbol{\beta}_g$ parameters and 95\% bootstrap CIs (Bulgaria). The reference category is in brackets.}
    \begin{tabular}{l r r | r r}
    \noalign{\smallskip}\hline\noalign{\smallskip}
     & $\hat{\boldsymbol{\beta}}_2$ & 95\% CI & $\hat{\boldsymbol{\beta}}_3$ & 95\% CI \\
         \noalign{\smallskip}\hline\noalign{\smallskip}
      Intercept & 5.99 & (1.14; 10.84) & 9.56 & (2.97; 16.16)  \\
      Age $30\vdash50$ ($<$30) & -1.02 & (-3.22; 1.16) & -1.45 & (-3.79; 0.89) \\
      Age $50\vdash65$ ($<$30) &  -2.75 & (-5.01; -0.50) & -3.79 & (-6.27; -1.30) \\
      Age $65\vdash75$ ($<$30) &  -3.77 & (-6.13; -1.41) & -4.67 & (-7.40; -1.95) \\
      Age $75+$ ($<$30) & -5.51 & (-8.08; -2.93) & -5.95 & (-8.85; -3.06) \\
      Health fair (bad) & 1.09 & (-0.19; 2.36) & 0.92 & (-0.61; 2.46)\\ 
      Health good (bad) & 1.26 & (0.05; 2.48) & 1.55 & (-0.09; 3.19) \\
      Hampered &  0.09 & (-0.91; 1.09) & -0.05 & (-1.37; 1.28) \\
      Born in country & -2.16 & (-5.46; 1.13) & -2.78 & (-6.46; 0.89) \\
      Male &  -0.57 & (-1.35; 0.21) & -0.65 & (-1.66; 0.36) \\
      Education low (high) & -1.52 & (-3.39; 0.36) & -3.64 & (-5.45; -1.82) \\ 
      Education medium (high) & -0.33 & (-1.86; 1.20) & -1.44 & (-2.76; -0.11) \\
      Partner education low (high) & -0.26 & (-2.29; 1.77) & -1.09 & (-3.44; 1.26) \\ 
      Partner education medium (high) & -0.31 & (-1.69; 1.07) & -0.95 & (-3.00; 1.10) \\
      Partner education not applicable (high) & -0.12 & (-1.78; 1.53) & -0.74 & (-3.13; 1.65) \\
      Income low (high) & -1.07 & (-2.32; 1.08) & -1.16 & (-2.95; 0.63) \\
      Income medium (high) & -0.45 & (-2.16; 1.59) & -0.47 & (-2.38; 1.43) \\
      Children & 0.36 & (-1.23; 1.95) & 0.36 & (-1.71; 2.43) \\
      Retired (employed) & -0.57 & (-2.50; 1.37) & -1.14 & (-3.26; 0.99) \\
      Unemployed (employed) & -0.18 & (-1.87; 1.52) & -0.57 & (-2.76; 1.62) \\     
      \noalign{\smallskip}\hline\noalign{\smallskip}
    \end{tabular}
    \label{beta_bulg}
\end{table}\\Table 14 shows the probability of belonging to each group for some covariates of interest. Once again, the probability of belonging to the most digitally skilled group is higher for young, healthy, and well-educated individuals, as well as for those with higher incomes. For example, the probability of belonging to the most digitalized group ($G$=3) ranges from 0.15 for individuals between 30 and 50 years of age to less than 0.01 for individuals aged 75 and over.
\begin{table}[ht!]
    \centering
     \caption{Predicted probabilities of belonging to each group for some covariates of interest (Bulgaria).}
     \setlength{\tabcolsep}{2pt}
    \footnotesize{
    \begin{tabular}{{lcccccccccccccccc}}
    \toprule
         &  \multicolumn{16}{c}{Variables}\\
        &  \multicolumn{5}{c}{Age} &  \multicolumn{3}{c}{Health}   &  \multicolumn{3}{c}{Education}   &  \multicolumn{3}{c}{Income}   &  \multicolumn{2}{c}{Children}\\
        \cmidrule(lr){2-6}
        \cmidrule(lr){7-9}
        \cmidrule(lr){10-12}
        \cmidrule(lr){13-15}
        \cmidrule(lr){16-17}
       Group  & $<30$ & $30\vdash50$ & $50\vdash65$ & $65\vdash75$ & $75+$ & Bad & Fair & Good & Low & Medium & High & Low & Medium & High & Yes & No \\
       \cmidrule(lr){1-1}
       \cmidrule(lr){2-6}
        \cmidrule(lr){7-9}
        \cmidrule(lr){10-12}
        \cmidrule(lr){13-15}
        \cmidrule(lr){16-17}
       $G$=1 & 0.00 & 0.63 & 0.92 & 0.97 & 0.99 & 0.68 & 0.15 & 0.11 & 0.80 & 0.51 & 0.00 & 0.54 & 0.42 & 0.05 & 0.44 & 0.10\\
       $G$=2 & 0.06 & 0.22 & 0.06 & 0.02 & 0.00 & 0.17 & 0.46 & 0.38 & 0.18 & 0.37 & 0.04 & 0.29 & 0.32 & 0.31 & 0.18 & 0.61 \\
         $G$=3 & 0.94 & 0.15 & 0.02 & 0.01 & 0.00 & 0.15 & 0.39 & 0.51 & 0.02 & 0.12 & 0.96 & 0.17 & 0.26 & 0.64 & 0.38 & 0.29 \\
         \bottomrule
    \end{tabular}}
    \label{p2.bulg}
\end{table}

\section{Conclusion}
\label{sec:5}
The paper exploits the Mixture of Latent Trait Analyzers (MLTA) with concomitant variables to cluster individuals in Finland, Italy and Bulgaria, according to their digitalization profiles, while also examining how individuals’ socio-economic and demographic characteristics, as well as intergenerational ties, influence such profiles. More specifically, data are represented as bipartite networks entailing the relations between individuals (sending nodes) and the digital skills (receiving nodes) they have. The analysis allows the identification of homogeneous groups of people, whose interpretation is very intuitive, since groups can be labeled according to their digital skill levels. Results highlight that individuals in the three analyzed countries show different levels of digitalization, with the highest in Finland and the lowest in Bulgaria. In both Italy and Bulgaria, MLTA identifies three patterns of digitalizations (clusters of individuals), while in Finland only two. This may be due to the generally higer level of digitalization in this country. However, also in Finland, the analyses of the role of covariates reveal important digital divides.\\In this respect, we find substantial differences in the level of digitalization according to age, health, income, and educational level. Furthermore, we note that Bulgaria is the country where the age digital divide is the strongest and, in fact, the probability of belonging to the most digitalized group is much lower for individuals over 30 years of age than in the other countries. Moreover, in this country, the probability of belonging to the least digitalized group varies from 63\% for individuals between 30 and 50 years of age to 100\% for individuals over 75 years of age. A similar consideration applies to education. In fact, the probability of belonging to the most digitalized group for those with a low or medium level of education is lower in Bulgaria than in Italy and Finland.\\Our results also show that the presence of children in the household seems to play an important role in the digitalization process in Finland and Italy. This may be due to the higher relevance of intergenerational ties in these countries with respect to Bulgaria. In fact, the mean predicted probability of belonging to the most digitalized group in Italy is 69\% for those who have children in the household versus 15\% for those who do not. In Finland, we observe a slightly lower gap between the two predicted probabilities (67\% versus 21\%).\\However, an important limitation regarding the estimated effect of intergenerational ties on digitalization is that the data we use do not provide information about ties between grandchildren and grandparents, which prevents the effect of intergenerational ties on clustering formation from  being accurately measured. Therefore, it might be possible that, by taking this aspect into account, the effect of intergenerational ties would also play a stronger role in Finland and Bulgaria too.\\Differently from most previous studies, the digital divides we highlight are not limited to crude measures such as {Internet} use but exploit more detailed information on digital skills.\\Our study also prompts to interesting avenues for future research. From a substantive point of view, future studies can address the estimation of the causal effect of intergenerational ties on the reduction of the digital divide between older adults and younger individuals.\\From a methodological point of view, the hierarchical structure of the ESS data (units nested within countries) suggests a possible extension of the MLTA to deal with unobserved sources of heterogeneity at multiple levels and to the possible correlation between the responses provided by units belonging to the same country. This will allow examining all ESS countries together.

\section*{Statements and Declarations}
A version of the manuscript is in production at Multivariate Behavioral Research.

\end{document}